\documentclass[aps,reprint,floats,prd,nofootinbib,superscriptaddress,10pt,longbibliography]{revtex4-1}

\usepackage[utf8]{inputenc}
\usepackage{amsmath}
\usepackage{graphicx}
\usepackage{dcolumn}
\usepackage{bm}
\usepackage{lipsum}
\usepackage{xcolor}
\usepackage{calc}
\usepackage{accents}
\usepackage{comment}
\usepackage{float}
\usepackage{diagbox}
\usepackage{ulem}
\usepackage{multirow}
\usepackage{hhline}
\usepackage{soul}
\usepackage{relsize,exscale}
\usepackage{array}
\usepackage{ctable}
\usepackage{cellspace} 

\newcolumntype{?}{!{\vrule width 0.12em}}

\def\hinvMpc{h\,{\rm Mpc}^{-1}}
\def\Mpcinvh{{\rm Mpc}/h}

\makeatletter
\newcommand\footnoteref[1]{\protected@xdef\@thefnmark{\ref{#1}}\@footnotemark}
\makeatother

\begin{document}


\title{Bayesian and frequentist investigation of prior effects in \\EFTofLSS analyses of full-shape BOSS and eBOSS data}

\author{Emil Brinch Holm}
\affiliation{Department of Physics and Astronomy, Aarhus University, DK-8000 Aarhus C, Denmark}
\author{Laura Herold}
\affiliation{Max-Planck-Institut f\"ur Astrophysik, Karl-Schwarzschild-Str.\ 1, 85748 Garching, Germany}
\author{Théo Simon}
\affiliation{Laboratoire Univers \& Particules de Montpellier (LUPM), CNRS \& Universit\'{e} de Montpellier (UMR-5299), Place Eug\`{e}ne Bataillon, F-34095 Montpellier Cedex 05, France}
\author{Elisa G. M. Ferreira}
\affiliation{Kavli Institute for the Physics and Mathematics of the Universe (WPI), UTIAS, The University of Tokyo, Chiba 277-8583, Japan}
\affiliation{Instituto de F\'isica, Universidade de S\~ao Paulo - C.P. 66318, CEP: 05315-970, S\~ao Paulo, Brazil}
\author{Steen Hannestad}
\affiliation{Department of Physics and Astronomy, Aarhus University, DK-8000 Aarhus C, Denmark}
\author{Vivian Poulin}
\affiliation{Laboratoire Univers \& Particules de Montpellier (LUPM), CNRS \& Universit\'{e} de Montpellier (UMR-5299), Place Eug\`{e}ne Bataillon, F-34095 Montpellier Cedex 05, France}
\author{Thomas Tram }
\affiliation{Department of Physics and Astronomy, Aarhus University, DK-8000 Aarhus C, Denmark}

\begin{abstract}
Previous studies based on Bayesian methods have shown that the constraints on cosmological parameters from the Baryonic Oscillation Spectroscopic Survey (BOSS) full-shape data using the Effective Field Theory of Large Scale Structure (EFTofLSS) depend on the choice of prior on the EFT nuisance parameters. In this work, we explore this prior dependence by adopting a frequentist approach based on the profile likelihood method, which is inherently independent of priors, considering data from BOSS, eBOSS and \textit{Planck}. 
We find that the priors on the EFT parameters in the Bayesian inference are informative and that prior volume effects are important. This is reflected in shifts of the posterior mean compared to the maximum likelihood estimate by up to $1.0\,\sigma$ ($1.6\,\sigma$) and in a widening of intervals informed from frequentist compared to Bayesian intervals by factors of up to $1.9$ ($1.6$) for BOSS (eBOSS) in the baseline configuration, while the constraints from \textit{Planck} are unchanged. 
Our frequentist confidence intervals give no indication of a tension between BOSS/eBOSS and \textit{Planck}. 
However, we find that the profile likelihood prefers extreme values of the EFT parameters, highlighting the importance of combining Bayesian and frequentist approaches for a fully nuanced cosmological inference. We show that the improved statistical power of future data will reconcile the constraints from frequentist and Bayesian inference using the EFTofLSS.

\end{abstract}

\maketitle

\section{Introduction}

\noindent In the last decades, the increasing precision of measurements of the cosmic microwave background (CMB) temperature fluctuations has reduced the experimental uncertainties to such an extent, that they are now dominated by cosmic variance~\cite{Planck:2018vyg}. 
This places an unavoidable limit on the amount of information extractable from the CMB and, therefore, additional cosmological probes are emerging, predominantly from large-scale structure (LSS) measurements. The Baryon Oscillation Spectroscopic survey (BOSS) of the Sloan Digital Sky survey \cite{BOSS:2012dmf}
is an example of a modern LSS probe, which will soon be joined by ambitious missions such as the Dark Energy Spectroscopic Instrument (DESI, \cite{DESI:2016fyo}), 
the Vera Rubin Observatory \cite{LSST:2008ijt} and the \textit{Euclid} space telescope \cite{Racca:2016qpi}, providing exciting new information about the LSS of the Universe. 

As the accuracy of the surveys increases, so does the demand for accurate theoretical model predictions. In particular, efficient computations of the statistics of inhomogeneities at small scales are crucial for drawing robust conclusions based on the upcoming data. $N$-body calculations, while giving accurate predictions, suffer from high demand for computational resources which usually make them unfeasible for full cosmological parameter inferences (although recent approaches based on machine learning may remedy this~\cite{Euclid:2020rfv, Fernandez:2022kxq, Lawrence_2010}). Instead, by compromising accuracy at the smallest scales, semi-analytic approaches based on perturbation theory (see \textit{e.g.}~\cite{Ivanov:2022mrd, Bernardeau:2001qr}, and references therein) may provide a computationally efficient alternative to $N$-body simulations. The recently developed effective field theory of large-scale structure (EFTofLSS) employs an effective field theory approach to predict the biased power spectrum up to mildly non-linear scales~\cite{Baumann:2010tm,Carrasco:2012cv,Senatore:2014via,Senatore:2014eva,Senatore:2014vja}.
The one-loop prediction of the EFTofLSS has allowed the determination of the $\Lambda$CDM parameters from the full-shape analysis of BOSS and eBOSS data at a precision higher than that from conventional baryon acoustic oscillation (BAO) and redshift-space distortion (RSD) analyses, and for some parameters even comparable to that of CMB experiments (see \textit{e.g.}, Refs.~\cite{DAmico:2019fhj,Ivanov:2019pdj,Colas:2019ret,DAmico:2020kxu,Chen:2021wdi,Zhang:2021yna,Zhang:2021uyp,Philcox:2021kcw,Simon:2022csv,Simon:2022lde,Chudaykin:2022nru,DAmico:2022gki}). 
Furthermore, the EFTofLSS may provide competitive and interesting constraints on models beyond $\Lambda$CDM (see \textit{e.g.}, Refs.~\cite{DAmico:2020tty,Simon:2022ftd,Kumar:2022vee,Nunes:2022bhn,Niedermann:2020qbw,Lague:2021frh,Carrilho:2022mon,Simon:2022adh,Smith:2022iax,Schoneberg:2023rnx}).

The EFTofLSS formalism is based on the most general parametrization of the evolution of the mildly non-linear scales admitted by symmetry. 
The coefficients of this parametrization, henceforth the \textit{EFT parameters}, although in theory obtainable from simulations are taken as free nuisance parameters in the statistical analyses. It was noted in Ref.~\cite{Carrilho:2022mon,Simon:2022lde} that this parameter structure may impact the results of Bayesian analyses through prior effects, especially when the data has weak constraining power. 
As a consequence, Ref.~\cite{Simon:2022lde} showed that different -- yet theoretically equivalent -- choices of the EFT parametrization result in discrepant Bayesian credible intervals and in point-estimate shifts sometimes on the order of $1\sigma$, particularly affecting the amplitude of matter fluctuations, $\sigma_8$. Additionally, Ref.~\cite{Carrilho:2022mon} found the priors on the EFT parameters to be informative and motivate a more comprehensive study of the effects of the parameter structure of the EFT sector. Ref.~\cite{Maus:2023rtr} argue that prior effects lead to 
a shift in $f\sigma_8$ in BOSS full-shape analyses based on an EFT implementation using the \texttt{Velocileptors} code~\cite{Chen:2020fxs,Chen:2020zjt,Chen:2021wdi},
partially explaining the difference with template fitting methods. Moreover, Refs.~\cite{Donald-McCann:2023kpx, Zhao:2023ebp} show that the use of a Jeffreys prior on the EFT parameters can mitigate biases in the standard EFT analysis.

Motivated by these previous results, in this paper, we complement the results of the standard Bayesian analysis with a profile likelihood analysis. The profile likelihood is a frequentist method based only on the maximum likelihood estimate (MLE) and, therefore, inherently reparametrization invariant and prior independent. 
Our goal is to understand the impact of priors on the EFT parameters on the inferred cosmological parameters and how this will change with more constraining data. 
In particular, we wish to answer the question: \textit{Does the seemingly low $\sigma_8$ value reconstructed from a Bayesian analysis of BOSS data under the EFTofLSS come from prior effects inherent to the Bayesian framework, rather than the true data likelihood?} Ultimately, our analysis demonstrates the importance of combining Bayesian and frequentist approaches for a fully nuanced inference from current and future LSS data.

This paper is structured as follows. In Sec.~\ref{sec:analysis_method}, we describe the respective analysis methods employed in the Bayesian and frequentist approaches and introduce the data sets used. In Sec.~\ref{sec:eftoflss}, we outline the EFTofLSS approach and give a detailed description of the two predominantly employed EFT parametrizations to be scrutinized. 
In Sec.~\ref{sec:impact}, we compare the two EFT parametrizations using the profile likelihood and contrast them to the MCMC results.
In Sec.~\ref{sec:prior_freq}, we study the influence of prior effects and discuss the issue that the EFT parameters take on extreme values in the frequentist setting.
In Sec.~\ref{sec:divby16}, we show that discrepancies between frequentist and Bayesian approaches subside with increasingly constraining data. 
Finally, we provide a profile likelihood analysis of the $\Lambda$CDM concordance model for the parameters $\sigma_8$, $h$, $\Omega_m$, $n_s$ and $\ln \left( 10^{10} A_s\right)$ with data from the BOSS and eBOSS surveys using the EFTofLSS formalism in Sec.~\ref{sec:results} and conclude in Sec.~\ref{sec:conclusions}.

\section{Analysis Methods}\label{sec:analysis_method}

\noindent The structure of the EFT parameters and their priors may impact the constraints on cosmological parameters derived from Bayesian inference. In particular, given a fixed choice of parametrization, we may classify the prior impact in terms of two separate effects, as was previously done in Ref.~\cite{Simon:2022lde}:
\begin{itemize}
    \item The \textit{prior weight effect}: Since the Bayesian posterior is proportional to the product of the prior and likelihood, non-flat priors will affect the posterior in a direct way when they do not align with the likelihood. This can manifest in, for example, a shift of the posterior peak or a scaling of its width.

    \item The \textit{prior volume effect}: Bayesian marginalization of the full-dimensional posterior involves integrating out the nuisance dimensions. Since in addition to the value of the posterior, an integral is sensitive to the volume in these directions, large parameter regions (of possibly non-maximal posterior values) are emphasized compared to smaller regions (of possibly larger posterior values).
\end{itemize}
Importantly, the volume effect can occur even with flat priors and is, therefore, an inescapable feature of the Bayesian method. Therefore, it becomes relevant to study the extent to which one's results are affected by volume effects. Since the profile likelihood is directly inferred from the likelihood, it is inherently independent of priors~\cite{pawitan} and is, therefore, an ideal tool for this. In Sec.~\ref{sec:prof}, we briefly review the use of profile likelihoods for inference, and in Sec.~\ref{sec:data} we describe our analysis pipeline.

\subsection{Profile Likelihood and Markov Chain Monte Carlo}\label{sec:prof}

\noindent The profile likelihood is a method in frequentist statistics, that allows to treat nuisance parameters (as opposed to marginalization, which is the commonly used method in Bayesian statistics). By splitting the full parameter space $\bm{\Theta}$ into two categories, ${\bm{\theta}}$ of $N$ parameters and ${\bm{\nu}}$ of $M$ (nuisance) parameters, the profile likelihood of ${\bm{\theta}}$ is obtained by maximization over all parameters in the complementary set of (nuisance) parameters ${\bm{\nu}}$ for fixed $\bm{\theta}$ ~\cite{pawitan},
\begin{equation}
    L({\bm{\theta}}) = \max_{\bm{\nu}}L({\bm{\theta}}, {\bm{\nu}}),
\end{equation}
where $L({\bm{\theta}}, {\bm{\nu}})$ represents the full likelihood function. Since the above is a MLE in the reduced parameter space ${\bm{\theta}}$, the profile likelihood is invariant under reparametrizations of the reduced parameter space ${\bm{\theta}}$~\cite{pawitan}. The reparametrization invariance of the profile likelihood will be particularly useful when comparing the different EFT parametrizations in Sec.~\ref{sec:impact}, which is more challenging with Bayesian methods since these can depend on the particular parametrization of the model and prior choices. In addition, the profile likelihood is inherently prior independent, thus automatically avoiding prior volume effects.

Frequentist methods like the profile likelihood are commonly used in particle physics but rarely used for cosmological inference. They recently gained more interest in the context of models beyond $\Lambda$CDM, which often contain many model parameters that are not well constrained by the data~\cite{Herold:2021ksg, Campeti:2022acx, Gomez-Valent:2022hkb, Campeti:2022vom, Reeves:2022aoi, Herold:2022iib, Holm:2022kkd, Cruz:2023cxy}, and in the context of efficient marginalization~\cite{Hadzhiyska:2023wae}.

To obtain parameter constraints in ${\bm{\theta}}$, we employ the Neyman construction, valid in the limit of a Gaussian likelihood of the data (also called the ``graphical construction'')~\cite{Neyman:1937uhy}: from the profile likelihood $L({\bm{\theta}}$), $\alpha$ confidence regions are given by the solution to $\Delta \chi^2 ({\bm{\theta}}) < F^{-1} (\alpha, N)$, where $F^{-1}$ is the inverse of the $\chi^2$ cumulative distribution function with $N$ degrees of freedom. For example, in the one-dimensional case ${\bm{\theta}}=\theta$, the $68 \% \ (95 \%)$ confidence intervals correspond to the values of $\theta$ for which $\Delta \chi^2 (\theta) < 1.0 \ (3.84)$. 
These confidence levels are exact when the likelihood is Gaussian, or, in the asymptotic limit of a large data set \cite{Wilks:1938dza}.
In this limit, the quantity $\Delta \chi^2 ({\bm{\theta}}) \equiv -2\log (L({\bm{\theta}} / L_\mathrm{max}))$ follows a $\chi^2$ distribution with $N$ degrees of freedom~\cite{pawitan} and the graphical method corresponds to the exact Neyman construction. Since for the BOSS and eBOSS data sets Gaussian likelihoods are employed, the graphical construction is exact, whereas parts of the \textit{Planck} likelihood are non-Gaussian \cite{Planck:2019nip} and we acknowledge that the graphical confidence intervals may be approximate. 
If the profile likelihood has a substantial overlap across a physical boundary of the parameter, an alternative Neyman construction needs to be used, also known as the Feldman-Cousins prescription~\cite{Feldman:1997qc}.
However, since the parameters studied in this work are well away from their physical boundaries, the Neyman construction is sufficient.  

Computing the profile likelihood amounts to optimizations in the reduced parameter space ${\bm{\nu}}$. Since evaluating the likelihood function $L({\bm{\theta}}, {\bm{\nu}})$ involves running the Einstein-Boltzmann solver, numerical gradients are noisy and inefficient~\cite{Planck:2013nga}. For the optimization, we therefore use \textit{simulated annealing}~\cite{Kirkpatrick:1983zz}, a gradient-free stochastic optimization algorithm (see \cite{Nygaard:2023cus} for efficient computation of profile likelihoods using an emulator and see \cite{Henrot-Versille:2016htt} for earlier approaches). 
The simulated annealing algorithm is based on chains with iteratively decreasing temperatures and step sizes, where the temperature $T>0$ modulates the likelihood function as $L ({\bm{\theta}}, {\bm{\nu}}) \rightarrow L({\bm{\theta}}, {\bm{\nu}})^{1/T}$.
Large temperatures smoothen the likelihood landscape, whereas small temperatures enhance peak structures. Thus, the chains are able to escape local optima while eventually being localized in a likelihood peak at low temperatures. Simulated annealing performs well against the noisy cosmological likelihood landscapes with many local optima~\cite{Hannestad:2000wx}, but may depend moderately on the particular temperature schedule employed. In practice, we inform the simulated annealing process with proposal covariance matrices and bestfits obtained from the corresponding MCMC analyses. Since the minimizations for each point in the profile are started from the global bestfit obtained from the MCMC, poor convergence would likely lead to an \textit{underestimation} of the width of the confidence interval, which would not have a strong impact on the conclusions in this paper as we find very large confidence intervals with the profile likelihood.
We ensure convergence and combat local optima by running each optimization several times. 
Due to the limited accuracy of the global bestfits caused by the finite sampling of the profile, we present the bestfit points in this paper as the optimum of the parabola fitted to the point of highest likelihood and its two neighboring points.
Our implementation of the simulated annealing algorithm\footnote{Publicly available at: \url{https://github.com/AarhusCosmology/montepython_public/tree/2211.01935}.} interfaces the \texttt{MontePython}~\cite{Audren:2012wb, Brinckmann:2018cvx} inference code with the Einstein-Boltzmann solver
\texttt{CLASS}~\cite{Blas:2011rf}\footnote{Publicly available at: \url{http://class-code.net}.}, which models the CMB coefficients and linear matter power spectra, and with \texttt{PyBird}~\cite{DAmico:2020kxu}\footnote{Publicly available at: \url{https://github.com/pierrexyz/pybird}.}, which models the full-shape of the galaxy power spectra from the EFTofLSS.
It is identical to the implementation used in Refs.~\cite{Holm:2022kkd, Cruz:2023cxy}. 

For all MCMCs performed in this study, we use the Metropolis-Hastings algorithm from \texttt{MontePython}, and we assume our MCMC chains to be converged with the Gelman-Rubin criterion $R - 1 < 0.05$.

In the following, we quote frequentist confidence intervals as the MLE $\pm$ $1\sigma$ obtained via the graphical Neyman method and we quote Bayesian credible intervals as the posterior mean $\pm$ $1\sigma$ obtained from the MCMC posterior. We will employ the following metric as a measure of the discrepancy between two approximately Gaussian posteriors or likelihoods,
\begin{align}
    \label{eq:sigma_distance}
    \sigma\text{-distance} \equiv \frac{|\theta_i - \theta_j |}{\sqrt{\sigma_{\theta, i}^2 + \sigma_{\theta, j}^2}}, 
\end{align}
where $\theta_i$ is the $i$'th point estimate of the parameter $\theta$ and $\sigma_{\theta, i}$ the corresponding standard deviation. The point estimates and standard deviations may be derived either from a posterior or from a profile likelihood. In the case that the two intervals are derived from the same model and the same statistical method (Bayesian or frequentist), but different data sets, the $\sigma$-distance coincides with the Gaussian tension metric employed, for example, in Ref.~\cite{Schoneberg_2019}.  When the point estimates are from different statistical paradigms, we instead normalize only by the Bayesian uncertainty,
\begin{align}
    \label{eq:sigma_distance_MCMC}
    \sigma\text{-distance} \equiv \frac{|\theta_\mathrm{Bayes.} - \theta_\mathrm{freq.} |}{\sigma_{\theta, \mathrm{Bayes.}}}, 
\end{align}
which can be interpreted as the significance of the bias between mean and MLE in units of the Bayesian error bars induced by the prior effects.

\subsection{Data sets and analysis choices}\label{sec:data}

\noindent In this paper we perform various MCMC and profile likelihood analyses using different datasets:

\begin{itemize}
    \item {\bf BOSS DR12 LRG}: 
    In our main analysis, we consider the BOSS luminous red galaxies data (LRG)~\cite{BOSS:2016wmc} (see Ref.~\cite{Reid:2015gra} for a description of the catalogues), with covariances built from the patchy mocks described in Ref.~\cite{Kitaura:2015uqa}. 
    The BOSS data are divided into four sky-cuts, corresponding to two galactic skies, denoted NGC and SGC, cut into to two redshift bins: LOWZ, which corresponds to the redshift range $0.2<z<0.43 \  (z_{\rm eff}=0.32)$, and CMASS, which corresponds to the redshift range $0.43<z<0.7  \ (z_{\rm eff}=0.57)$.
    For LOWZ we analyse the galaxy power spectrum up to $k_{\rm max} = 0.20 \hinvMpc$, while for CMASS we analyse it up to $k_{\rm max} = 0.23 \hinvMpc$.
    In this study, we use the EFT likelihood of the full shape of the BOSS LRG power spectrum pre-reconstructed multipoles, including the monopole and the quadrupole, 
    measured and described in Ref.~\cite{Zhang:2021yna} and referred to as ``BOSS''. 
    We also consider ``BOSS+BAO'', which additionally includes the cross-correlation of the pre-reconstructed measurements with post-reconstruction BAO compressed parameters obtained in Ref.~\cite{DAmico:2020kxu} on the post-reconstructed power spectrum measurements of Ref.~\cite{Gil-Marin:2015nqa}.
    
    \item {\bf eBOSS DR16 QSO}: 
    We also consider the quasars (QSO) data from the extended Baryon Oscillation Spectroscopic Survey (eBOSS)~\cite{eBOSS:2020yzd} (see Ref.~\cite{Ross:2020lqz} for a description of the catalogues), with covariances built from the EZmocks described in Ref.~\cite{Chuang:2014vfa}.
    The eBOSS data are divided into two sky-cuts, corresponding to two galactic skies, denoted NGC and SGC, in the redshift range $0.8 < z < 2.2 \ (z_{\rm eff}=1.52)$. 
    We analyse the eBOSS QSO galaxy power spectrum up to $k_{\rm max} = 0.24 \hinvMpc$.
    In this study, we use the EFT likelihood of the full shape of the eBOSS QSO power spectrum pre-reconstructed multipoles from Ref.~\cite{Simon:2022csv} and the measurements of Ref.~\cite{Beutler:2021eqq}, including the monopole and the quadrupole, 
    which is referred to as ``eBOSS''. 

    \item{\bf BBN likelihood}: As in Ref.~\cite{Simon:2022lde}, unless specified otherwise, we impose a Gaussian likelihood on $\omega_b \sim \mathcal{N}(0.02268, 0.00038)$, where $\mathcal{N}(\bar{x}, \sigma_x)$ denotes a Gaussian centered on $\bar{x}$ with standard deviation $\sigma_x$, coming from BBN experiments~\cite{Schoneberg_2019}. This likelihood is based on the theoretical prediction of \cite{Consiglio_2018}, the experimental Helium fraction of~\cite{Aver_2015} and the experimental Deuterium fraction of~\cite{Cooke_2018}.

    \item {\bf \textit{Planck}}: Finally, we compare the BOSS and eBOSS results with the low-$l$ CMB TT, EE, and the high-$l$ TT, TE, EE data, as well as the gravitational lensing potential reconstruction from {\it Planck} 2018~\cite{Planck:2018vyg}, referred to as ``\textit{Planck}''.
\end{itemize}

\noindent For the BOSS and eBOSS analyses, we vary five cosmological parameters: 
\begin{equation} \label{eq:cosmo_params}
 \lbrace \omega_\mathrm{cdm}, \, \omega_b, \, h, \, \ln (10^{10} A_s), \, n_s \rbrace , 
\end{equation}
corresponding to the physical cold dark matter and baryon energy density, the reduced Hubble constant, the $\log$-amplitude of the primordial fluctuations and the scalar spectral index, respectively.\footnote{For runs that include \textit{Planck} data, we also vary $\tau_{\rm reio}$, the re-ionization optical depth, within a large flat prior.}
For the MCMC, we assume large flat priors, and for the profile likelihood, we scan a parameter range that covers at least the $95 \%$ confidence interval.
For the LSS data, unless specified otherwise, we always include the BBN likelihood mentioned above.
To facilitate comparison with previous studies, we present our cosmological results on $\lbrace \sigma_8, \ h, \ \Omega_m, \ n_s, \ \ln\left( 10^{10} A_s\right) \rbrace$, corresponding respectively to the clustering amplitude, the reduced Hubble constant, the fractional matter abundance as well as the scalar spectral index and amplitude of primordial fluctuations from~\eqref{eq:cosmo_params}.
Finally, for all analyses performed we use the {\it Planck} convention for the neutrinos, namely we take two massless and one massive species with $m_{\nu} = 0.06$ eV~\cite{Planck:2018vyg}.

\section{The effective field theory of large-scale structure formalism}\label{sec:eftoflss}

\noindent To model the full shape of the BOSS and eBOSS power spectra, we use the EFTofLSS theoretical prediction at one-loop order. In the literature, several prescriptions have been proposed for the EFT parameters. In line with Refs.~\cite{Nishimichi:2020tvu, Simon:2022lde}, we consider the two most commonly used parametrizations, namely the ``West coast'' (WC) parametrization, the one used in the \texttt{PyBird}~\cite{DAmico:2020kxu} likelihood, and the ``East coast'' (EC) parametrization, the one used in the \texttt{CLASS-PT}~\cite{Chudaykin:2020aoj, Philcox:2021kcw} likelihood\footnote{Let us note that there exists another EFT likelihood implemented in the public code \texttt{Velocileptors}~\cite{Chen:2020fxs,Chen:2020zjt,Chen:2021wdi}, with different prior choices on the EFT parameters.}.
In this section, we describe these two EFT parametrizations and the associated priors.

\subsection{Power spectrum at one-loop order}

\noindent In this study, we use the monopoles $P_0(z,k)$ and quadrupoles $P_2(z,k)$ of the BOSS LRG and eBOSS QSO power spectra given by: 
\begin{equation}
    P_\ell(z,k)=\frac{2\ell+1}{2}\int^1_{-1}d\mu \, \mathcal{L}_\ell(\mu)P_{g}(z,k,\mu) \, ,
\end{equation}
where $\mathcal{L}_\ell$ corresponds to the Legendre polynomial of order $\ell$, and $\mu = \hat{z}\cdot\hat{k}$ is the angle between the line-of-sight ${\boldsymbol z}$ and the wavevector of the Fourier mode ${\boldsymbol k}$. $P_{g}(z,k,\mu)$ corresponds to the EFTofLSS power spectrum of biased tracers in redshift space at one-loop order,\footnote{The first formulation of the EFTofLSS was carried out in Eulerian space in Refs.~\cite{Carrasco:2012cv,Baumann:2010tm} and in Lagrangian space in \cite{Porto:2013qua}. Once this theoretical framework was established, many efforts were made to improve this theory and make it predictive, such as the understanding of renormalization \cite{Pajer:2013jj, Abolhasani:2015mra}, the IR-resummation of the long displacement fields \cite{Senatore:2014vja, Baldauf:2015xfa, Senatore:2014via, Senatore:2017pbn, Lewandowski:2018ywf, Blas:2016sfa}, and the computation of the two-loop matter power spectrum \cite{Carrasco:2013sva, Carrasco:2013mua}. Then, this theory was developed in the framework of biased tracers (such as galaxies and quasars) in Refs. \cite{Senatore:2014eva, Mirbabayi:2014zca, Angulo:2015eqa, Fujita:2016dne, Perko:2016puo, Nadler:2017qto}.} which reads, within the WC parametrization~\cite{Perko:2016puo}: 
\begin{widetext}
\begin{align} \label{eq:power_spectrum}
& P_{g}(k,\mu) =  Z_1(\mu)^2 P_{11}(k)  + 2 Z_1(\mu) P_{11}(k)\left( c_\text{ct}\frac{k^2}{{ k^2_\textsc{m}}} + c_{r,1}\mu^2 \frac{k^2}{k^2_\textsc{r}} + c_{r,2}\mu^4 \frac{k^2}{k^2_\textsc{r}} \right) \\  
& + 2 \int \frac{d^3q}{(2\pi)^3}\; Z_2({\bf q},{\bf k}-{\bf q},\mu)^2 P_{11}(|{\bf k}-{\bf q}|)P_{11}(q) + 6 Z_1(\mu) P_{11}(k) \int\, \frac{d^3 q}{(2\pi)^3}\; Z_3({\bf q},-{\bf q},{\bf k},\mu) P_{11}(q) \nonumber \\
& + \frac{1}{\Bar{n}_g}\left( c_{\epsilon,0} +  c_{\epsilon}^{\textrm{mono}} \frac{k^2}{k^2_\textsc{m}} + 3c_{\epsilon}^{\textrm{quad}} \left(\mu^2-\frac{1}{3}\right) \frac{k^2}{k^2_\textsc{m}}  \right), \nonumber
\end{align}
\end{widetext}
where $f$ is the growth factor, and $P_{11}(k)$ is the linear matter power spectrum (calculated with the \texttt{CLASS} code). 
In the following, we give a description of the different terms of Eq.~\eqref{eq:power_spectrum}:
\begin{itemize}
    \item The first term corresponds to the linear galaxy power spectrum in redshift space, also known as the Kaiser formula \cite{Kaiser:1987qv}. This term depends on $b_1(z)$, which is the linear galaxy bias parameter [see Eq.~\eqref{eq:z1}].
    
    \item The second term proportional to $k^2 Z_1(\mu) P_{11}(k)$ corresponds to the contribution of the one loop-order counterterms. $c_\text{ct}$ is a linear combination of the dark matter sound speed~\cite{Baumann:2010tm,Carrasco:2012cv} and a higher-derivative bias~\cite{Senatore:2014eva}, while $c_{r,1}$ and $c_{r,2}$ represent the redshift-space counterterms~\cite{Senatore:2014vja}. Let us note that in this analysis, we do not consider $c_{r,2}$ (which belongs to a $\mu^4-$term), since we do not include the hexadecapole. Without the latter, this term is degenerate with $c_{r,1}$.

    \item The second line corresponds to the one-loop perturbation contribution, which depends on four galaxy bias parameters appearing in Eqs.~\eqref{eq:z1}-\eqref{eq:z3}: $b_i$, with $i=[1,4]$.

    \item Finally, the last line, inversely proportional to the mean galaxy number density $\Bar{n}_g$, corresponds to the stochastic contribution, which depends on three stochastic terms: $c_{\epsilon,0}$, $c_{\epsilon}^{\textrm{mono}}$ and $c_{\epsilon}^{\textrm{quad}}$. The first term describes a constant shot noise, while the other two terms correspond to the scale-dependant stochastic contributions of the monopole and the quadrupole.\\
\end{itemize}

\noindent In the contributions of the one loop-order counterterms and the stochastic terms there are two scales that govern the EFT expansions: $k_\textsc{m}^{-1}$, corresponding to the spatial extension of the observed objects~\cite{Senatore:2014eva}, and $k_\textsc{r}^{-1}$, corresponding to the ``dispersion'' scale~\cite{Senatore:2014vja}. While the former controls the spatial derivative expansion, the latter is the scale that renormalizes the velocity products appearing in the redshift-space expansion.

In Eq.~\eqref{eq:power_spectrum}, $Z_1$, $Z_2$, and $Z_3$, corresponding to the redshift-space galaxy density kernels of order $n$, are given by~\cite{Perko:2016puo}:
\begin{widetext}
\begin{align} \label{eq:z1}
    Z_1({\bf q}_1) & = K_1({\bf q}_1) +f\mu_1^2 G_1({\bf q}_1) = b_1 + f\mu_1^2  \, ,\\ \label{eq:z2}
    Z_2({\bf q}_1,{\bf q}_2,\mu) & = K_2({\bf q}_1,{\bf q}_2) +f\mu_{12}^2 G_2({\bf q}_1,{\bf q}_2)+ \, \frac{1}{2}f \mu q \left( \frac{\mu_2}{q_2}G_1({\bf q}_2) Z_1({\bf q}_1) + \text{perm.} \right) \, ,\\ \label{eq:z3}
    Z_3({\bf q}_1,{\bf q}_2,{\bf q}_3,\mu) & = K_3({\bf q}_1,{\bf q}_2,{\bf q}_3) + f\mu_{123}^2 G_3({\bf q}_1,{\bf q}_2,{\bf q}_3) \\ \nonumber
    &+ \frac{1}{3}f\mu q \left(\frac{\mu_3}{q_3} G_1({\bf q}_3) Z_2({\bf q}_1,{\bf q}_2,\mu_{123}) +\frac{\mu_{23}}{q_{23}}G_2({\bf q}_2,{\bf q}_3)Z_1({\bf q}_1)+ \text{cyc.}\right)  \, , 
\end{align}
where \vspace*{-0.75cm}
\begin{align} \label{eq:k1}
     K_1 & = b_1 \, , \\ \label{eq:k2}
     K_2({\bf q}_1,{\bf q}_2) & = b_1 \frac{{\bf q}_1\cdot {\bf q}_2 (q_1^2 + q_2^2)}{2 q_1^2 q_2^2}+ b_2\left( F_2({\bf q}_1,{\bf q}_2) -  \frac{{\bf q}_1\cdot {\bf q}_2 (q_1^2 + q_2^2)}{2 q_1^2 q_2^2} \right) + b_4 \, , \\ \label{eq:k3}
     K_3({\bf q},-{\bf q},{\bf k}) & = \frac{b_1}{504 k^3 q^3}\left( -38 k^5q + 48 k^3 q^3 - 18 kq^5 + 9 (k^2-q^2)^3\log \left[\frac{k-q}{k+q}\right] \right) \nonumber \\
    &+ \frac{b_3}{756 k^3 q^5} \left( 2kq(k^2+q^2)(3k^4-14k^2q^2+3q^4)+3(k^2-q^2)^4 \log \left[\frac{k-q}{k+q}\right]  \right) \nonumber \\
    & +\frac{b_1}{36 k^3 q^3} \left( 6k^5 q + 16 k^3 q^3 - 6 k q^5 + 3 (k^2 - q^2)^3 \log \left[\frac{k-q}{k+q}\right] \right) \, ,
\end{align}
\end{widetext}
with $\mu= {\bf q} \cdot \hat{{\bf z}}/q$, ${\bf q} = {\bf q}_1 + \dots +{\bf q}_n$, and $\mu_{i_1\ldots  i_n} = {\bf q}_{i_1\ldots  i_n} \cdot \hat{{\bf z}}/q_{i_1\ldots  i_n}$, ${\bf q}_{i_1 \dots i_m}={\bf q}_{i_1} + \dots +{\bf q}_{i_m}$. 
In Eqs.~\eqref{eq:z1}-\eqref{eq:z3}, $G_i$ represents the \textit{velocity kernels} of the standard perturbation theory, and $K_i$ represents the \textit{galaxy density kernels}, defined as in Eqs.~\eqref{eq:k1}-\eqref{eq:k3}~\cite{Senatore:2014eva,Angulo:2015eqa,Fujita:2016dne}, where $F_2$ is the symmetrized second-order density kernel from the standard perturbation theory~\cite{Bernardeau:2001qr}.

\subsection{Different parametrizations}\label{sec:parameterizations}

\subsubsection{WC parametrization}

\noindent In the previous section, we expressed the power spectrum in the framework of the WC parametrization using 10 EFT terms: 4 bias
parameters ($b_i$, with $i=[1,4]$), 3 counterterms ($c_\text{ct}$, $c_{r,1}$ and $c_{r,2}$), and 3 stochastic terms ($c_{\epsilon,0}$, $c_{\epsilon}^{\textrm{mono}}$ and $c_{\epsilon}^{\textrm{quad}}$). 
In this study, we set to zero \cite{DAmico:2019fhj} the parameters $c_{r,2}$ (degenerated with $c_{r,1}$, as we do not include the hexadecapole), implying that we end up with 9 EFT parameters for each sky-cut of the BOSS LRG and eBOSS QSO data. In the \texttt{PyBird} likelihood, instead of using $b_2$ and $b_4$, we use linear combinations of these parameters: $c_2 = (b_2+b_4)/\sqrt{2}$ and $c_4 = (b_2-b_4)/\sqrt{2}$. Given that $b_2$ and $b_4$ are almost completely anti-correlated (at $\sim 99\%$ according to Ref.~\cite{DAmico:2019fhj}), 
the standard procedure is to set $c_4 =0$. In addition, $c_{\epsilon}^{\textrm{mono}}$ is also set to 0 in the \texttt{PyBird} baseline analysis since the functions that are multiplied by this parameter were found to be small compared to 
the signal-to-noise ratio associated with the BOSS volume \cite{Schmittfull:2020trd,DAmico:2019fhj}. In this study, we include $c_4$ and $c_{\epsilon}^{\textrm{mono}}$ as free parameters in our analysis when comparing the WC parametrization with the EC parametrization in Sec.~\ref{sec:impact}, which ensures mathematical equivalence between the EC and WC parametrizations. On the other hand, for our cosmological results (where we only use the WC parametrization) we adopt the standard \texttt{PyBird} convention and set $c_4 = c_{\epsilon}^{\textrm{mono}} = 0$ to facilitate easier comparison with previous works. In Sec.~\ref{sec:impact}, we find that fixing or freeing $c_4$ and $c_{\epsilon}^{\textrm{mono}}$ changes the frequentist confidence intervals for $\sigma_8$, indicating that the effect of these two EFT parameters is not negligible.

Within the WC parametrization, we set $k_{\rm M} = 0.7 h \, {\rm Mpc}^{-1}$, $k_{\rm R} = 0.35 h \, {\rm Mpc}^{-1}$ and $\bar{n}_g = 4 \cdot 10^{-4} ({\rm Mpc}/h)^3$ for the BOSS LRG data~\cite{DAmico:2021ymi}, and $k_{\rm M} = 0.7 h \, {\rm Mpc}^{-1}$, $k_{\rm R} = 0.25 h \, {\rm Mpc}^{-1}$ and $\bar{n}_g = 2 \cdot 10^{-5} ({\rm Mpc}/h)^3$ for the eBOSS QSO data~\cite{Simon:2022csv} in Eq.~\eqref{eq:power_spectrum}.

\subsubsection{EC parametrization}

\noindent We now turn to the EC parametrization which is used by the \texttt{CLASS-PT} likelihood~\cite{Chudaykin:2020aoj}. In the following, we list the differences between the two parametrizations, and comment on how to switch from one to the other:
\begin{itemize}
    \item {\bf Bias parameters}: the EC parametrization uses the $\lbrace \tilde b_1, \tilde b_2, b_{\mathcal{G}_2}, b_{\Gamma_3} \rbrace$ basis~\cite{Mirbabayi:2014zca}, which is related to the previous basis $\lbrace b_1, b_2, b_3, b_4 \rbrace$ in the following way~\cite{Fujita:2020xtd}: 
    \begin{align}
        b_1 &= \tilde b_1, \nonumber \\
        b_2 &= \tilde b_1 + \frac{7}{2}b_{\mathcal{G}_2}, \nonumber \\
        b_3 &= \tilde b_1 + 15 b_{\mathcal{G}_2} + 6  b_{\Gamma_3}, \nonumber \\
    b_4 &= \frac{1}{2}\tilde b_2 - \frac{7}{2}b_{\mathcal{G}_2}\, .         \label{eq:basis}
    \end{align}
    These two bases are equivalent and describe the one-loop contribution.

    \item {\bf Counterterms}: in the EC parametrization, the definition of the counterterms $\lbrace c_{0}, c_{2}, c_{4} \rbrace$ changes slightly with respect to the WC parametrization $\lbrace c_{ct}, c_{r,1},c_{r,2} \rbrace$: $k_{\rm M}$ and $k_{\rm R}$ are now absorbed in the counterterm coefficients, such that $c_{0} \propto c_{ct}/k_{\rm M}^2$, $c_{2} \propto c_{r,1}/k_{\rm R}^2$ and $c_{4} \propto c_{r,2}/k_{\rm R}^2$. Note that in the EC parametrization, these counterterms are not unitless. In this analysis, we fix $c_{4}=0$ as we do not include the hexadecapole.

    \item {\bf Stochastic terms}: we use the same definition for the stochastic parameters as for the WC parametrization. Further, the EC parametrization uses $k_{\rm M} = 0.45 \hinvMpc$ and $\bar n \simeq 3 \cdot 10^{-4} (\Mpcinvh)^3$.
\end{itemize}

\begin{table*}[tb]
\begin{tabular}{?l?c|c?c|c?}
\specialrule{.12em}{0em}{0em}
                      & \multicolumn{2}{|c|}{WC Priors}   & \multicolumn{2}{|c|}{EC Priors} \\
\hline
Parameter type        & Parameter    & MCMC prior         &  Parameter              & MCMC prior \\
\hline
\multirow{4}{*}{Bias} & $b_{1}$      & flat $[0,4]$       & $\tilde b_{1}$          & flat $[0,4]$ \\
                      & $c_{2}$      & flat $[-4,4]$      & $\tilde b_{2}$          & $\mathcal{N}(0, 1)$\\
                      & $c_{4}\ (*)$ & flat $[-4,4]$      & $b_{\mathcal{G}_2}$     & $\mathcal{N}(0, 1)$\\
                      & $b_{3}$      & $\mathcal{N}(0,2)$ & $b_{\Gamma_3}$          & $\mathcal{N}(\tfrac{23}{42}(b_1-1),1)$\\
\hline
\multirow{2}{*}{Counterterms} 
                      & $c_{ct}$     & $\mathcal{N}(0,2)$ & $c_{0}/[{\rm Mpc}/h]^2$ & $\mathcal{N}(0, 30)$ \\
                      & $c_{r,1}$    & $\mathcal{N}(0,2)$ & $c_{2}/[{\rm Mpc}/h]^2$ & $\mathcal{N}(30, 30)$\\
\hline
\multirow{2}{*}{Stochastic} 
 & $c_{\epsilon,0}$                  & $\mathcal{N}(0,2)$ & $c_{\epsilon,0}$ & $\mathcal{N}(0, 2)$\\
 & $c_{\epsilon}^\textrm{mono}\ (*)$ & $\mathcal{N}(0,2)$ & $c_{\epsilon}^{\textrm{mono}}$ & $\mathcal{N}(0, 2)$  \\
 & $c_{\epsilon}^\textrm{quad}$      & $\mathcal{N}(0,2)$ & $c_{\epsilon}^{\textrm{quad}}$ & $\mathcal{N}(0, 2)$ \\
\specialrule{.12em}{0em}{0em}
\end{tabular}
\caption{Standard priors on the EFT parameters in the WC and EC parametrizations used for MCMC analyses in this paper. In the WC parametrization, $b_1$ and $c_2$ vary within flat priors, whereas in the EC parametrization, $\tilde b_1$ varies within a flat prior, $\tilde b_2$ and $b_{\mathcal{G}_2}$ vary within Gaussian priors, while Gaussian priors are imposed on the other parameters before analytically marginalizing them. 
In the profile likelihood analyses, we mimic the case without priors by multiplying all priors by a factor 100.
The two parameters with $(*)$ are set to 0 for our cosmological results, but we include them for the comparison with the EC parametrization in Sec.~\ref{sec:impact} to ensure perfect equivalence between the two parametrizations.
$\mathcal{N}(\Bar{x}, \sigma_x)$ corresponds to a Gaussian prior on the parameter $x$ with a mean value of $\Bar{x}$ and a standard deviation of $\sigma_x$.}
\label{tab:wc_ec_priors}
\end{table*}

\noindent Note that the EC baseline parametrization includes a next-to-next leading order parameter, $\tilde c$, in front of a term in $\sim k^4 P_{11}(k)$. In order to be consistent with the WC parametrization, we do not include this term in this analysis, which implies that we end up with 9 EFT parameters that are equivalent to the WC ones.

In this paper, in line with Ref.~\cite{Simon:2022lde}, the results of the EC parametrization are obtained with \texttt{PyBird}, which supports both the EC and WC parametrizations. 
This facilitates exploration of the differences in the inferred cosmological parameters introduced by the priors and parametrizations of the EFT parameters without the need to take into account differences in data and codes, namely the different implementations in \texttt{CLASS-PT} and \texttt{PyBird} (we invite the interested reader to refer to Ref.~\cite{Simon:2022lde} for such a comparison).

\subsection{Priors}

\noindent In the left half of Tab.~\ref{tab:wc_ec_priors}, we summarize the MCMC standard priors used for the 9 parameters in the \texttt{PyBird} code. In general, given the perturbative nature of the theory, the one-loop contribution should be smaller than the tree-level contribution. The latter is given by the Kaiser formula, which depends on the linear bias $b_1$, implying that the other EFT parameters should be in $\sim \mathcal{O}(b_1)$.
In the standard WC analysis, i.e. $c_4 = c_{\epsilon}^{\textrm{mono}} =0$, the parameters $b_1$ and $c_2$ vary within flat priors, while the other EFT parameters, \textit{i.e.}, those which enter linearly into Eq.~\eqref{eq:power_spectrum}, are analytically marginalized with Gaussian priors following the procedure of App.~C of Ref.~\cite{DAmico:2020kxu}. 

In the right half of Tab.~\ref{tab:wc_ec_priors}, we summarize the MCMC standard priors used for the 9 parameters in the \texttt{CLASS-PT} likelihood. The main difference to the WC priors is that the EC priors are mainly based on simulations~\cite{Ivanov:2021kcd}. In the standard EC analysis, $\tilde b_1$ varies within a flat prior, and $\tilde b_2$ and $b_{\mathcal{G}_2}$ vary within Gaussian priors, while the other EFT parameters are analytically marginalized within Gaussian priors. 

For the profile likelihood analysis, in theory, we do not need to include priors. However, for practical reasons related to the implementation of the EFT likelihood, we mimic the case without priors by multiplying the bounds of the flat priors and the standard deviation of the Gaussian priors in Tab.~\ref{tab:wc_ec_priors} by 100. In App.~\ref{ap:A} we check that this leads to an effectively flat prior. Lastly, we refrain from applying the analytical marginalization from appendix~C of Ref.~\cite{DAmico:2019fhj}, commonly used in the standard analysis.
Instead, we use the analytical approximation (without marginalization) from the same reference to estimate, at each point in the optimizations, the best-fitting values of the EFT parameters that have Gaussian priors in the standard configuration, having checked explicitly that this approximation works to good precision even with flat priors.

\section{Consistency of EFTofLSS from profile likelihood analyses}\label{sec:tech}

\noindent In this section, we compare the two EFTofLSS parametrizations introduced in Sec.~\ref{sec:parameterizations}, contrast them to the standard MCMC results, explore the impact of the Bayesian priors, and illustrate explicitly the effect of more constraining data.
We take the example of the amplitude of matter clustering\footnote{Note that the definition of $\sigma_8$, which is in units of Mpc/$h$, depends also on the background cosmology and, therefore, alternative measures of the amplitude of matter fluctuations have been proposed~\cite{Sanchez:2020vvb, Semenaite:2021aen, Semenaite:2022unt, Brieden:2021edu}.}, $\sigma_8$, which was found to be particularly affected by prior effects~\cite{Carrilho:2022mon,Simon:2022lde}.

\subsection{EC vs.\ WC parametrizations and comparison to MCMC}\label{sec:impact}

\begin{figure}[tb]
    \includegraphics[width=\columnwidth]{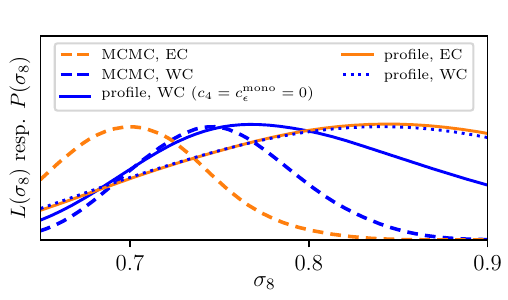}
    \caption{\label{fig:wc_vs_ec} Marginalized MCMC posteriors (dashed) and profile likelihoods (solid) of $\sigma_8$ within the WC (blue) and EC parametrizations (orange), for BOSS+BAO data. The two statistical approaches and two parametrizations yield different intervals for $\sigma_8$. If $c_4$ and $c_\epsilon^\mathrm{mono}$ are allowed to vary in the WC parametrization, the MCMC posteriors do not agree (dashed lines), while the WC-profile likelihood (blue dotted) agrees with the EC-profile likelihood (orange solid), confirming that the two mathematically equivalent parametrizations lead to the same likelihood. In the remainder of the paper, we adopt the WC-standard convention ($c_4=c_\epsilon^\mathrm{mono}=0$, blue solid).}
\end{figure}

\noindent In Fig.~\ref{fig:wc_vs_ec}, we compare the one-dimensional marginalized MCMC posteriors $P(\sigma_8)$ to the profile likelihoods $L(\sigma_8)$, which are normalized by their individual MLEs. We use BOSS full-shape data combined with reconstructed BAO data based on the WC (blue) and EC (orange) parametrizations, respectively. We find that the Bayesian MCMC posteriors differ from the frequentist profile likelihoods in both WC and EC parametrizations, respectively, indicating that priors and/or marginalization have an impact on the constraints on $\sigma_8$ in the Bayesian analysis, as was already pointed out in Ref.~\cite{Simon:2022lde}. 

In the WC parametrization, the standard configuration includes setting $c_4=c_\epsilon^\mathrm{mono}=0$. Mathematically, the WC parametrization is only equivalent to the EC parametrization if $c_4$ and $c_\epsilon^\mathrm{mono}$ are taken as free parameters (see Sec.~\ref{sec:parameterizations}). However, even if $c_4$ and $c_\epsilon^\mathrm{mono}$ are free to vary, the MCMC posteriors in the two parametrizations (dashed lines), using the recommended standard priors in Tab.~\ref{tab:wc_ec_priors}, do \textit{not} yield the same credible interval:
\begin{equation}
    \begin{split}
        &\sigma_8 = 0.748^{+0.043}_{-0.048}  \quad\quad \text{(MCMC, WC)},  \\
        &\sigma_8 = 0.700 \pm 0.044          \quad \text{(MCMC, EC)}. 
    \end{split}    
\end{equation}
Ref.~\cite{Simon:2022lde} showed that this difference, which corresponds to a $\sigma$-distance of $0.7\,\sigma$ (as defined in Eq.~\ref{eq:sigma_distance}), can be attributed to the different prior configurations in the WC and EC parametrizations (and not to differences in the implementation of the codes). 

The profile likelihoods, on the other hand, do not depend on priors, since they are constructed solely from the MLE, and are reparametrization invariant.
Therefore, two profile likelihoods from the same data set will agree if the underlying models are equivalent, i.e., if the range of their possible predictions coincide.
We explicitly confirm that if $c_4$ and $c_\epsilon^\mathrm{mono}$ are free to vary, the profile likelihood in the WC parametrization (blue dotted) agrees with the profile likelihood in the EC parametrization (orange solid) up to numerical accuracy:
\begin{equation}
    \begin{split}
        &\sigma_8 = 0.850 \pm 0.119  \quad \text{(profile, WC)},  \\
        &\sigma_8 = 0.850 \pm 0.117  \quad \text{(profile, EC)}. 
    \end{split}    
\end{equation}
Note that in Fig.~\ref{fig:wc_vs_ec}, we show the individually normalized profiles, but we checked that the absolute values of the likelihood at each point are also approximately equal with maximum differences of $\Delta\chi^2<0.2$, which can be attributed to uncertainties in the optimization.
This consistency check at the example of $\sigma_8$ confirms the mathematical equivalence of the WC and EC parametrizations. 

Since the recommended standard configuration in the WC parametrization includes setting $c_4=c_\epsilon^\mathrm{mono}=0$, we use this as the baseline setting for both Bayesian and frequentist analyses in the remainder of the paper to facilitate comparison with previous work. The profile likelihood in the baseline configuration (blue solid line in Fig.~\ref{fig:wc_vs_ec}, $c_4=c_\epsilon^\mathrm{mono}=0$) yields:
\begin{equation}
    \sigma_8 = 0.7699 \pm 0.0851 \quad \text{(profile, WC-base)},
\end{equation}
which differs from the profile likelihood with free $c_4$, $c_\epsilon^\mathrm{mono}$ in the WC parametrization (blue dotted) by $0.6\,\sigma$. Fixing $c_4$ and $c_\epsilon^\mathrm{mono}$ also leads to a reduction of the width of the frequentist confidence interval by $30\%$. This indicates that $c_4$ and $c_\epsilon^\mathrm{mono}$ have an impact on the inference for $\sigma_8$, which cannot be neglected for the profile likelihood analysis. 
Explicitly checking the bestfit values of these two EFT parameters close to the global MLE, i.e. the minimum of the profile likelihood, reveals that these parameters take on non-zero values as large as $c_4\approx 57$ and $c_\epsilon^\mathrm{mono}\approx 38$ (depending on the particular skycut), pointing to an important role played by these two parameters and motivating closer inspection of the impact of analysis choices regarding the EFT parameters, which we present in the next section.

\subsection{Role of EFT ``priors'' in the frequentist setting} \label{sec:prior_freq}
\noindent It is instructive to look at the values attained by the EFT parameters in the 
frequentist framework, which requires varying all parameters in very large flat ranges. Let us recall that the EFT parameters in the WC parametrization should be of order unity in order to conserve the perturbative nature of the EFTofLSS~\cite{DAmico:2021ymi}. Yet, we 
find that they take on extreme values at most points in the profile. For example, Fig.~\ref{fig:nonmarg_params} in App.~\ref{ap:A} shows the values of the EFT parameters at each point in the $\sigma_8$ profile with the baseline configuration (WC, $c_4=c_\epsilon^\mathrm{mono}=0$), which finds values like $b_3 \approx 26$ and $c_\text{ct} \approx 23$. Similarly large values appear in the $\sigma_8$ profile using the EC configuration, where we find as large values as $b_2 \approx 53$ and $b_{\mathcal{G}_3} \approx 38$. This indicates that the profile likelihood includes parts of the EFT parameter space in the analysis in which the EFT prediction is no longer valid. In the Bayesian analysis this issue is addressed by imposing narrow Gaussian priors on the EFT parameters (see Tab.~\ref{tab:wc_ec_priors}). However, as we will now show, imposing a specific (subjective) prior has a direct impact on the inferred uncertainty in $\sigma_8$.

Indeed, the intervals from the profile likelihoods in Fig.~\ref{fig:wc_vs_ec} are broader than the intervals from the MCMC posteriors by factors of 2.6 to 2.7 (for $c_4$, $c_\epsilon^\mathrm{mono}$ free). To explore whether this significant loss in constraining power can be explained by the information content of the priors in the Bayesian analysis, we construct a profile likelihood subject to the same ``priors'' as the Bayesian analysis: If the non-flat Bayesian priors were well-founded, they could in principle be promoted to likelihoods, be interpreted as genuine data, and thus used in the profile likelihood construction. 

\begin{figure}[tb]
    \includegraphics[width=0.5\textwidth]{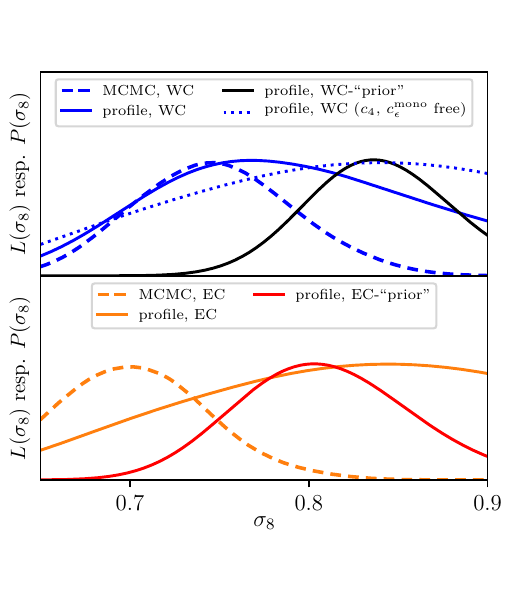}
    \caption{\label{fig:prior_effects_WC_EC} Same as Fig.~\ref{fig:wc_vs_ec} but including profile likelihoods with Gaussian data likelihoods on the EFT parameters, which correspond to the standard WC (top, black line) and EC priors (bottom, red line). The Gaussian likelihoods lead to a reduction of the width of the profiles almost to the level of the MCMC posterior and to small shifts of the MLE. However, the posterior and profile do still not overlap, which can be explained by prior volume effects in the Bayesian inference.}
\end{figure}

In Fig.~\ref{fig:prior_effects_WC_EC}, we show the impact of including Gaussian likelihoods on the EFT parameters, which correspond to the standard priors in the WC (top, black solid line, with free $c_4$, $c_\epsilon^\mathrm{mono}$) and EC parametrization (bottom, red solid line), as quoted in Tab.~\ref{tab:wc_ec_priors}.
Including the Gaussian data likelihoods gives the following frequentist confidence intervals:
\begin{equation}
    \begin{split}
        &\sigma_8 = 0.817 \pm 0.049  \quad \text{(profile, WC-``priors'')},  \\
        &\sigma_8 = 0.783 \pm 0.060  \quad \text{(profile, EC-``priors'')}. 
    \end{split}    
\end{equation}
We observe a strong increase in constraining power, reducing the width of the frequentist intervals almost to the level of the Bayesian intervals, indicating that the priors on the EFT parameters are informative. We also observe a slight shift in the global MLE toward the mean 
of the posterior as a result of including the Gaussian likelihoods on the EFT parameters. However, the shift thus introduced is not enough to reconcile the frequentist and Bayesian results; we observe a $\sigma$-distance of about $1\,\sigma$ for both the WC and EC parametrizations. This is an indication that there is not only a prior \textit{weight} effect, which is a direct result of the multiplication of the prior, but also a prior \textit{volume} effect, which is a result of the marginalization (see Sec.~\ref{sec:analysis_method}) of some of the model parameters.
This is in agreement with Ref.~\cite{Maus:2023rtr}, which finds similar results for $f\sigma_8$ using a profile likelihood analysis based on \texttt{Velocileptors}~\cite{Chen:2020fxs,Chen:2020zjt,Chen:2021wdi} (see \textit{e.g.}\ their Fig.~3). 
Moreover, Ref.~\cite{Donald-McCann:2023kpx} find that the posteriors of several EFT parameters, e.g. $c_4$, $c_\epsilon^\mathrm{mono}$, $b_3$, $c_\mathrm{ct}$ among others, are dominated by the prior information (see their Fig.~8), reinforcing our conclusions that the priors on the EFT parameters are informative.
In App.~\ref{ap:A}, we go one step further and illustrate the impact of changing the prior width on the profile likelihood of $\sigma_8$.

We conclude this section with the observation that both statistical approaches come with disadvantages in the context of BOSS+BAO data. While the results of the Bayesian analysis depend on informative (subjective) priors and are influenced by volume effects, the frequentist analysis takes into account parts of the EFT parameter space in which the theory is no longer valid, which reflects a significant loss of constraining power. As a way forward, we explore the impact of using more constraining data than the BOSS+BAO data in the next section.

\subsection{Effect of more constraining data}\label{sec:divby16}

\noindent In the asymptotic limit of infinite data, the likelihood will dominate the Bayesian prior, and prior effects will vanish accordingly~\cite{pawitan}. Consequently, Bayesian and frequentist constraints will converge to the same answer as the model is better constrained by data. 

\begin{figure}[tb]
    \includegraphics[width=\columnwidth]{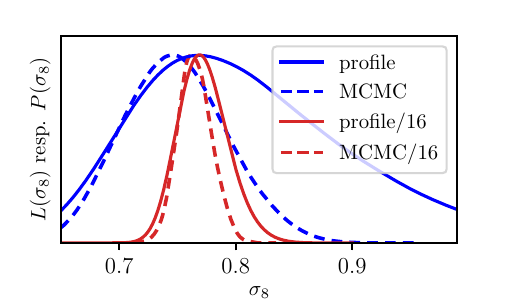}
    \caption{\label{fig:divby16} Profile likelihoods (solid) and marginalized MCMC posteriors (dashed) of $\sigma_8$ in the WC parametrization under BOSS+BAO data (blue) and the same data but with a data covariance divided by $16$ (red). This illustrates how more constraining power reduces the difference between the Bayesian and frequentist approaches.}
\end{figure}

To illustrate this point, we rescaled the BOSS covariance matrix by a factor $16$, simulating a prospective situation with less uncertainties or, equivalently, a larger data volume, roughly corresponding to that of future galaxy surveys such as DESI \cite{Aghamousa:2016zmz} or Euclid \cite{Amendola:2016saw}. 
In Fig.~\ref{fig:divby16}, we compare the constraints on $\sigma_8$ from the rescaled data covariance to those obtained from the unscaled data covariance using both MCMC and profile likelihoods, normalized to their MLE. Note that from now on, we show only results in the WC parametrization, using the default configuration $c_4=c_\epsilon^\mathrm{mono}=0$.
The constraints on $\sigma_8$ as well as the $\sigma$-distances, as defined in Eq.~\eqref{eq:sigma_distance_MCMC}, are given in Table~\ref{tab:divby16}.

\begin{table}[b]
\begin{tabular}{?Sl|Sc|Sc?}
\specialrule{.12em}{0em}{0em}
& BOSS + BAO & BOSS$/16$ + BAO \\
\hline
MCMC (mean $\pm 1\sigma$) & $0.748\pm 0.045$ & $0.765\pm 0.015$ \\ 
\hline
profile (bf. $\pm 1\sigma$) & $0.770 \pm 0.085$ & $0.770 \pm 0.018$ \\ 
\hline
$\sigma$-distance & $0.49\sigma$ & $0.33\sigma$ \\
\specialrule{.12em}{0em}{0em}
\end{tabular}
\caption{Constraints on $\sigma_8$ from the marginalized MCMC posteriors and profile likelihoods of Fig.~\ref{fig:divby16}. The last row gives the $\sigma$-distances between the MCMC/profile constraints.}
\label{tab:divby16}
\end{table}

With the reduced data covariance, the profile and posterior are narrower and roughly centered around the same value of $\sigma_8$. When reducing the data covariance, the posterior mean value obtained from the MCMC moves closer to the MLE (i.e., the maximum of the profile likelihood), while the MLE is unchanged since the case with reduced data covariance is based on the same power-spectra data. 
Table~\ref{tab:divby16} shows that the consistency improves from $0.49\sigma$ to $0.33\sigma$ when we reduce the data covariance.

This improved consistency between the bestfit and the posterior mean of the MCMC shows that the prior influence decreases as the data volume increases, as already pointed out in Ref.~\cite{Simon:2022lde}.
Thus, discrepancies between Bayesian and frequentist methods can be seen as due to a lack of data, which will improve as more data is obtained in the future. Furthermore, one may hope that more data will aid in constraining the EFT parameters helping to avoid extreme values at which the EFT is no longer valid, though this is not guaranteed. Hence, we can look to future galaxy surveys to improve the situation for EFTofLSS analyses using either statistical method.


\section{Profile likelihood results on cosmological parameters \label{sec:results}}

\noindent In this section, we present profile likelihood results from the EFTofLSS applied to BOSS, eBOSS and \textit{Planck} data for five selected $\Lambda$CDM parameters, $\sigma_8$, $h$, $\Omega_m$, $n_s$, and $A_s$, and compare to the credible intervals from the Bayesian MCMC. While lacking more constraining data, comparison of frequentist and Bayesian methods can help to gain a more nuanced view of the data. For both frequentist and Bayesian setups we use the standard WC parametrization (setting $c_4=c_\epsilon^\mathrm{mono}=0$) of the \texttt{PyBird} likelihood and for the MCMC the default prior configuration from Ref.~\cite{DAmico:2019fhj} as above. 
\begin{figure*}[t]
    \includegraphics[width=\textwidth]{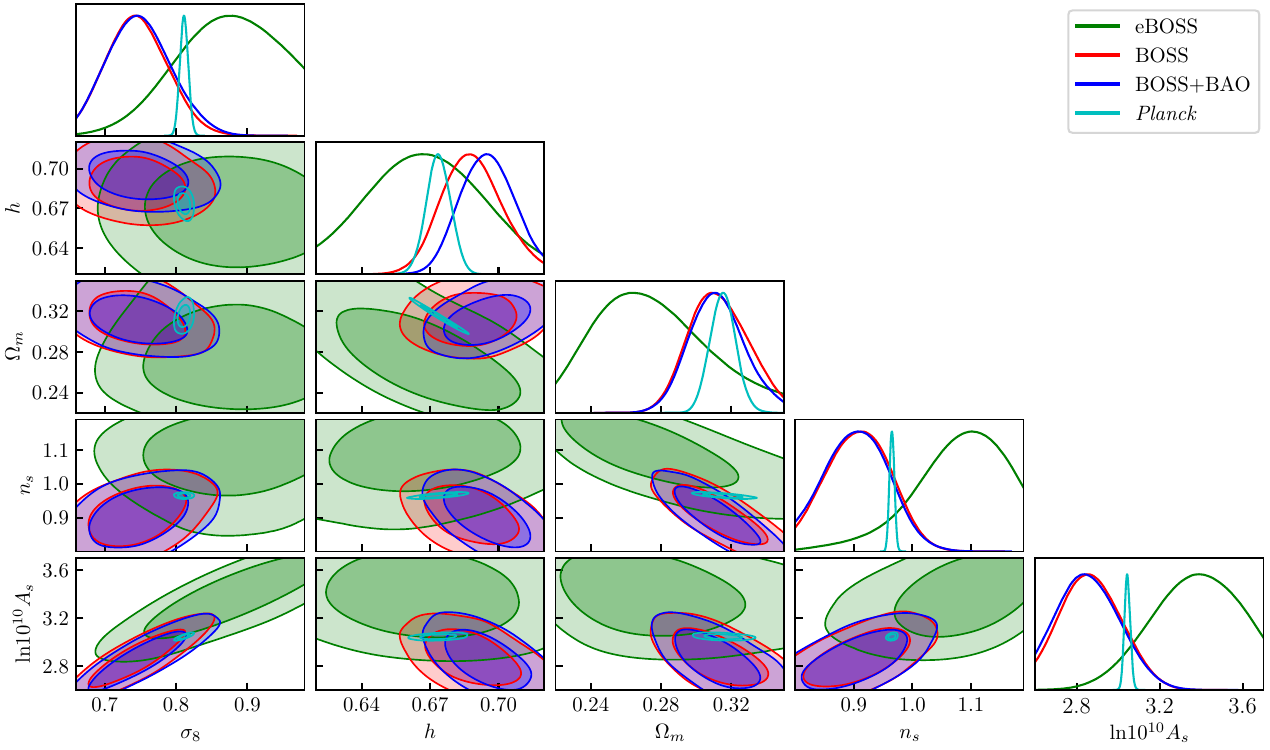}
    \caption{\label{fig:mcmc} MCMC posteriors for five selected $\Lambda$CDM parameters using four different data sets, described in Sec.~\ref{sec:data}.}
\end{figure*}

\textbf{Bayesian results.} Firstly, Fig.~\ref{fig:mcmc} shows the one-dimensional marginalized posterior distributions and the $68\%$ and $95\%$ two-dimensional marginalized posteriors obtained from our MCMC analyses for the BOSS, BOSS + BAO, eBOSS, and \textit{Planck} data (see Sec.~\ref{sec:data} for details). The general picture, which corroborates previous results using the WC parametrization of the EFTofLSS~\cite{DAmico:2019fhj, Simon:2022lde}, is that the parameter constraints from BOSS and eBOSS show overall agreement with \textit{Planck} data up to $1.6\,\sigma$. All $\sigma$-distances, as defined in Eq.~\ref{eq:sigma_distance}, are summarized in Table \ref{tab:sigma-distances}. 
We confirm that BOSS+BAO data prefers slightly lower values of $\sigma_8$ than \textit{Planck} data at a significance of $1.4\,\sigma$. Note that this difference is larger in the EC parametrization corresponding to a $\sigma$-distance of $2.5\,\sigma$ (see Sec.~\ref{sec:impact}). Moreover, we find that BOSS+BAO data prefers slightly larger values of $h$ than \textit{Planck} at a significance of $1.6\,\sigma$ and eBOSS prefer slightly larger values of $n_s$ and $A_s$ than \textit{Planck} at a significance of $1.4\,\sigma$ to $1.5\,\sigma$, while having a weaker constraining power compared to BOSS data. The inclusion of the reconstructed BAO data does not alter the constraints from BOSS significantly, the most significant being a $0.4\sigma$ shift on $h$.\footnote{Compared to previous analyses, especially Ref.~\cite{Simon:2022csv}, we do not set $n_s$ to the \textit{Planck} value, which explains why our LSS constraints are somewhat weaker and why we have a stronger inconsistency between eBOSS and BOSS.}
 
\begin{figure*}[t]
    \includegraphics[width=\textwidth]{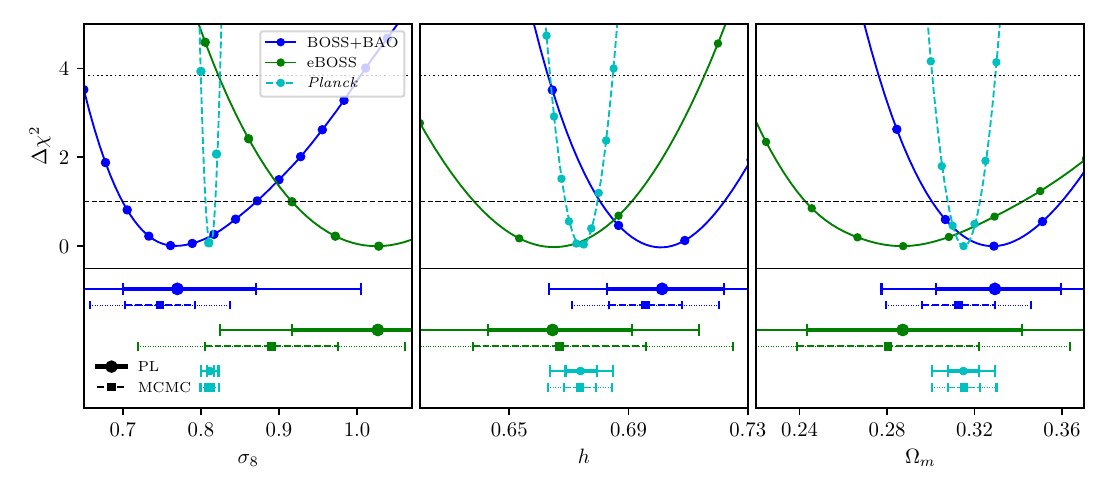}
    \includegraphics[width=0.67\textwidth]{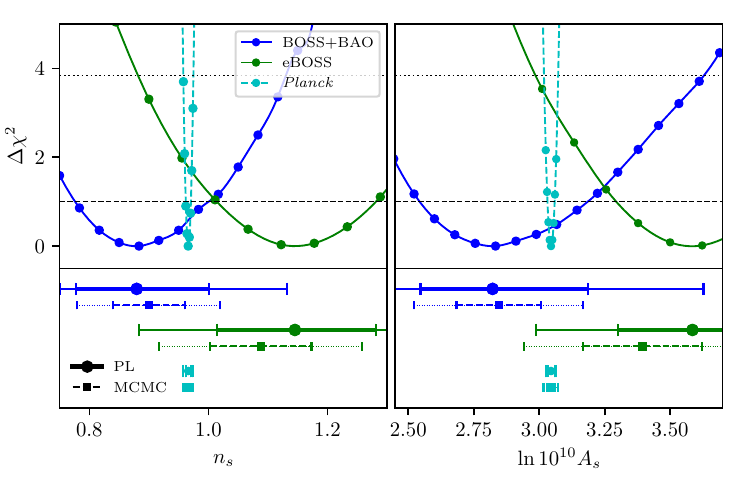}
    \caption{\label{fig:profiles} 
    Profile likelihoods for five selected $\Lambda$CDM parameters using the three main data sets described in Sec.~\ref{sec:data}. For each of the parameters, the top subplots show the profile likelihoods in terms of the quantity $\Delta \chi^2 (\theta) = -2\log(L(\theta)/L_\mathrm{max})$, where $L_\mathrm{max}$ is the MLE. The bottom subplots show the $68\%$ and $95\%$ confidence intervals derived from the profiles (solid) as well as the $68\%$ and $95\%$ credible intervals obtained from the Bayesian analysis (dashed) of Fig.~\ref{fig:mcmc}. The profile constraints differ from the MCMC constraints for BOSS+BAO and eBOSS data, while the \textit{Planck} constraints are roughly unchanged. We find no indication for a tension between any of the considered data sets.}
\end{figure*}

\textbf{Frequentist results.} Fig.~\ref{fig:profiles} shows the profile likelihood results for the cosmological parameters $\sigma_8$, $h$, $\Omega_m$, $n_s$, and $A_s$. For each of the parameters, the top panels show the profile likelihoods in terms of the $\Delta \chi^2$, such that according to the Neyman construction for a Gaussian likelihood the intersections with $\Delta \chi^2 = 1 \ (3.84)$, shown as the dashed (dotted) horizontal line, gives the $68\%$ ($95 \%$) confidence interval. The bottom panels show such constructed confidence intervals, along with the corresponding credible intervals obtained from the MCMC analyses. Note that the confidence intervals for \textit{Planck} have been constructed from fitting the $\Delta \chi^2$ to a parabola, which is the fit shown in the figure. This is appropriate since the $\Lambda$CDM profiles are Gaussian under \textit{Planck} data~\cite{Planck:2013nga}.
For a visual comparison, individual profiles and posteriors for each parameter and data combination can be found in Fig.~\ref{fig:app_comp} of App.~\ref{ap:full_results}. Our constraints are summarized in Table~\ref{tab:results}, and the global bestfitting parameters in the BOSS+BAO and eBOSS data sets are given in App.~\ref{ap:bestfits}. 
In Tab.~\ref{tab:sigma-distances}, we indicate the $\sigma$-distances between several combinations of experiments for either the MCMC or the profiles, while in Tab.~\ref{tab:sigma-dist_bf_mean}, we display the $\sigma$-distances between posterior mean and MLE for each data set.
In the following, we will discuss the profile results and compare them to the MCMC results for each data set individually.

\textbf{BOSS \& the ``$\sigma_8$-discrepancy''.} Our profile likelihood confidence intervals for the BOSS+BAO data are in good agreement with the confidence intervals from \textit{Planck} data for all five cosmological parameters at less than $1.4\, \sigma$ and we find no indication for a tension. Removing the reconstructed BAO data leads only to sub-$\sigma$ shifts, the largest being in $h$, which is  $0.7\sigma$ larger when including the reconstructed BAO data (as is the case for the MCMC analysis). When comparing to the credible intervals from the MCMC, the most striking feature is that the confidence intervals from the profile are much wider, \textit{e.g.},\ the $68 \%$ profile confidence intervals are wider by a factor of 1.4 to 1.9 than the MCMC credible intervals. As already discussed in Sec.~\ref{sec:prior_freq}, this cannot fully be attributed to prior volume effects, and is consequently an indication that the priors on the EFT parameters in the Bayesian approach are informative and lead to tighter constraints on the cosmological parameters. 
The point estimates of profile and MCMC differ only slightly; we find $\sigma$-distances between posterior mean and MLE, as defined in Eq.~\ref{eq:sigma_distance_MCMC}, up to $1\sigma$, namely $\sim 0.5\sigma$ on $h$ and $\sigma_8$, and $\sim 1\sigma$ on $\Omega_m$ (see Tab.~\ref{tab:sigma-dist_bf_mean}).
As discussed in Sec.~\ref{sec:prior_freq}, note that in our BOSS and BOSS+BAO results, we observe that the EFT parameters take on extreme values, which reflects in considerably larger uncertainties and questions the validity of the EFTofLSS in our profile likelihood analysis.

Our results corroborate previous findings \cite{Donald-McCann:2023kpx,Simon:2022lde,Simon:2022csv} that there is no indication for a ``$\sigma_8$ discrepancy'' between BOSS and \textit{Planck} data. While in the Bayesian analysis the $\sigma$-distance between $\sigma_8$ posteriors of BOSS+BAO data based on the WC (EC) parametrization and \textit{Planck} is $1.4\,\sigma$ ($2.5\,\sigma$), this is reduced to $0.49\,\sigma$ ($0.33\,\sigma$) for the profile. This reduction of the $\sigma$-distance is mainly due to the increase of the errorbar by a factor of 1.9 (2.7) along with a shift of the MLE compared to the posterior mean to slightly larger values of $\sigma_8$. These results suggest treating the somewhat curious $2.5\,\sigma$ discrepancy in $\sigma_8$ obtained in the MCMC analysis using the EC parametrization cautiously since it depends on the EC convention of the EFT parameter priors and on prior-volume effects inherent to the Bayesian framework. 

\begin{table*}[t]
\begin{tabular}{?Sc|Sc|Sc|Sc|Sc|Sc|Sc?} 
 \specialrule{.12em}{0em}{0em}
 & & $\sigma_8$ & $h$ & $\Omega_m$ & $n_s$ & $\ln 10^{10}A_s$ \\ 
\hhline{|=|=|=|=|=|=|=|}
\multirow{2}{*}{BOSS} & PL & $0.8025\pm 0.0925$ & $0.6816\pm 0.0209$ & $0.3197\pm 0.0291$ & $0.9499\pm 0.1349$ & $3.0304 \pm 0.3167$  \\
\cline{2-7}
 & MCMC & $0.7443 \pm 0.0433$ & $0.6889 \pm 0.0136$ & $0.3137 \pm 0.0174$ & $0.9050\pm 0.0576$ & $2.8610\pm 0.1543$ \\
  \hline
 \multirow{2}{*}{BOSS$+$BAO rec.} & PL & $0.7699 \pm 0.0851$ & $0.7013\pm 0.0183$ & $0.3293\pm 0.0281$ & $0.8795\pm 0.1078$ & $2.8222\pm 0.2918$ \\
 \cline{2-7}
 & MCMC & $0.7476\pm 0.0450$ & $0.6957\pm 0.0123$ & $0.3126\pm 0.0170$ & $0.8997\pm 0.0602$ & $2.8455\pm 0.1612$ \\
\hline
\multirow{2}{*}{eBOSS} & PL & $1.0267 \pm 0.1179$ & $0.6645 \pm 0.0233$ & $0.2872\pm 0.0490$ & $1.1454 \pm 0.1326$ & $3.5852\pm 0.3065$ \\
\cline{2-7}
 & MCMC & $0.8903\pm 0.0856$ & $0.6668\pm 0.0291$ & $0.2804\pm 0.0416$ & $1.0880\pm 0.0853$ & $3.3940\pm 0.2266$ \\
 \hline
 \multirow{2}{*}{\textit{Planck}} & PL & $0.8122 \pm 0.0063$ & $0.6742\pm 0.0054$ & $0.3151\pm 0.0074$ & $0.9663\pm 0.0044$ & $3.0453\pm 0.0139$ \\
 \cline{2-7}
 & MCMC & $0.8112\pm 0.0058$ & $0.6737\pm 0.0054$ & $0.3153\pm0.0074$ & $0.9651\pm 0.0042$ & $3.0446\pm 0.0142$ \\
 \specialrule{.12em}{0em}{0em}
 \end{tabular}
 \caption{\label{tab:results} $68\%$ C.L. constraints obtained in this paper. Profile likelihood (PL) constraints represent the bestfit and confidence interval from the Neyman construction described in Sec.~\ref{sec:prof}; the quantity in $\pm$ is the average of the absolute difference between the lower and upper bounds and the bestfit (noting that the profiles are largely Gaussian). The MCMC constraints represent the mean of the marginalized one-dimensional posterior and its associated $68\%$ credible interval.}
\vspace{0.2cm}
\begin{tabular}{?Sc|Sc|Sc|Sc|Sc|Sc|Sc?} 
 \specialrule{.12em}{0em}{0em}
 & & \hspace*{0.5cm}$\sigma_8$\hspace*{0.5cm} & \hspace*{0.5cm}$h$\hspace*{0.5cm} & \hspace*{0.4cm}$\Omega_m$\hspace*{0.4cm} & \hspace*{0.5cm}$n_s$\hspace*{0.5cm} & $\ln 10^{10}A_s$ \\ 
\hhline{|=|=|=|=|=|=|=|}
\multirow{2}{*}{BOSS+BAO vs.\ \textit{Planck}} & PL & $0.49\sigma$ & $1.33\sigma$ & $0.48\sigma$ & $0.78\sigma$ & $0.70\sigma$  \\
\cline{2-7}
 & MCMC & $1.40\sigma$ & $1.63\sigma$ & $0.15\sigma$ & $1.08\sigma$ & $1.23\sigma$ \\
  \hline
 \multirow{2}{*}{eBOSS vs.\ \textit{Planck}} & PL & $1.82\sigma$ & $0.39\sigma$ & $0.56\sigma$ & $1.34\sigma$ & $1.76\sigma$ \\
 \cline{2-7}
 & MCMC & $0.92\sigma$ & $0.23\sigma$ & $0.83\sigma$ & $1.44\sigma$ & $1.54\sigma$ \\
\hline
\multirow{2}{*}{BOSS+BAO vs.\ eBOSS} & PL & $1.77\sigma$ & $1.18\sigma$ & $0.74\sigma$ & $1.53\sigma$ & $1.72\sigma$ \\
\cline{2-7}
 & MCMC & $1.48\sigma$ & $0.91\sigma$ & $0.72\sigma$ & $1.80\sigma$ & $1.87\sigma$ \\
 \specialrule{.12em}{0em}{0em}
 \end{tabular}
 \caption{\label{tab:sigma-distances} $\sigma$-distances, as defined in Eq.~\ref{eq:sigma_distance}, for five selected parameters between different data sets.}
\vspace{0.2cm}
\begin{tabular}{?Sc|Sc|Sc|Sc|Sc|Sc?} 
 \specialrule{.12em}{0em}{0em}
& \hspace*{0.5cm}$\sigma_8$\hspace*{0.5cm} & \hspace*{0.5cm}$h$\hspace*{0.5cm} & \hspace*{0.4cm}$\Omega_m$\hspace*{0.4cm} & \hspace*{0.5cm}$n_s$\hspace*{0.5cm} & $\ln 10^{10}A_s$ \\
\hhline{|=|=|=|=|=|=|}
BOSS+BAO & $0.50\sigma$ & $0.46\sigma$ & $0.98\sigma$ & $0.34\sigma$ & $0.14\sigma$  \\
\hline
 eBOSS & $1.59\sigma$ & $0.08\sigma$ & $0.16\sigma$ & $0.67\sigma$ & $0.84\sigma$ \\
\hline
\textit{Planck} & $0.16\sigma$ & $0.08\sigma$ & $0.03\sigma$ & $0.29\sigma$ & $0.05\sigma$ \\
 \specialrule{.12em}{0em}{0em}
 \end{tabular}
 \caption{\label{tab:sigma-dist_bf_mean} Distance between posterior mean and bestfit in units of the standard deviation, $\sigma$, of the posterior, as defined in Eq.~\ref{eq:sigma_distance_MCMC}.}
\end{table*}

\textbf{eBOSS.} The profile likelihood confidence intervals from eBOSS data show mild discrepancies with \textit{Planck} and BOSS+BAO data for some parameters, \textit{e.g.},\ $\sigma_8$ is $1.82\,\sigma$ ($1.77\,\sigma$) higher than for \textit{Planck} (BOSS+BAO) and $\ln10^{10}A_s$ is $1.82\,\sigma$ ($1.72\,\sigma$) higher than for \textit{Planck} (BOSS+BAO), which is similar to the MCMC analyses (see Tab.~\ref{tab:sigma-distances}). Otherwise, the parameter constraints of eBOSS are within around $\lesssim 1.5\,\sigma$ of the constraints from \textit{Planck} and BOSS+BAO. When comparing to the MCMC constraints, we find that the width of the $68\%$ confidence intervals of the profile is a factor 1.2 to 1.6 wider than the credible intervals of the MCMC. The bestfit obtained from the profile is within $1\sigma$ of the posterior mean obtained from the MCMC except for the parameter $\sigma_8$, where the bestfit is at a $1.59\,\sigma$ higher value than the posterior mean. However, as with BOSS data, we also find extreme values of the EFT parameters under eBOSS data.

\textbf{\textit{Planck}.} For comparison, we also constructed profile likelihoods for \textit{Planck} data. We find very good agreement between the constraints from profile likelihoods and MCMC for \textit{Planck} data. The width of the confidence and credible intervals agree within less than $8\%$ and the shifts between bestfit and posterior mean are less than $0.3\,\sigma$. This corroborates the results in Ref.~\cite{Planck:2013nga}, which used \textit{Planck} 2013 intermediate results and also found very good agreement between both methods. The good agreement between the profile likelihood and MCMC are expected due to the high constraining power of \textit{Planck} data, which dominates over any prior information.
We note that for all cosmological parameters, the \textit{Planck} constraints are in-between the BOSS and eBOSS ones, indicating no tension between the CMB and the galaxy clustering data.

\section{Conclusions}\label{sec:conclusions}

\noindent Motivated by previous Bayesian studies that found a prior dependence of the inferred cosmological parameters from BOSS full-shape data using the EFTofLSS \cite{Carrilho:2022mon, Simon:2022lde, Donald-McCann:2023kpx}, in this work, we present frequentist profile likelihood constraints to view this matter from a different statistical point of view. In particular, two of the commonly used parametrizations of the EFTofLSS, the WC~\cite{DAmico:2020kxu} and EC parametrizations~\cite{Chudaykin:2020aoj}, give different constraints on the cosmological parameters of up to $\sim1\,\sigma$ in a Bayesian analysis \cite{Simon:2022lde}. 

Using the profile likelihood, we find that the WC and EC parametrizations yield the same confidence interval for $\sigma_8$, confirming that the two parametrizations are mathematically equivalent, i.e.,\ they describe the same space of model predictions for the galaxy power spectrum multipoles (see Fig.~\ref{fig:wc_vs_ec} in Sec.~\ref{sec:impact}).\footnote{This equivalence requires the free variation of two EFT parameters in the WC parametrization ($c_4$ and $c_\epsilon^\mathrm{mono}$, see Sec.~\ref{sec:eftoflss}), which are typically fixed to zero in the standard WC convention. Instead, we find a strong correlation between these parameters and $\sigma_8$, motivating further study.}
However, we find that the profile likelihood gives constraints on $\sigma_8$ that are factors of $>2$ wider than the constraints based on the MCMC posterior. 
Moreover, we observed that several of the EFT parameters take on extreme values during the profile likelihood analysis, indicating that the frequentist analysis takes into account parts of the EFT parameter space beyond the intended use of the theory, in which the perturbative nature might be broken. 
This issue is addressed in the Bayesian case by imposing narrow Gaussian priors on the EFT parameters. 
If these priors were well founded, \textit{e.g.},\ motivated from theory, simulations, or other observations, the priors could in principle be promoted to data likelihoods in the frequentist analysis. Although the priors on the EFT parameters are not rigorously motivated, we explore the effect of including Gaussian data likelihoods in the frequentist analysis, which correspond to the priors in the Bayesian analysis. 
We find that the inclusion of the Gaussian likelihoods on the EFT parameters reduces the width of the constraints almost to the level of the ones inferred from the MCMC posterior and keeps the EFT parameters in the intended range (see Fig.~\ref{fig:prior_effects_WC_EC} in Sec.~\ref{sec:prior_freq}). However, it also leads to a shift of the confidence interval of $\sigma_8$. This demonstrates that the priors on the EFT parameters in the Bayesian analysis are informative and influence the inferred cosmological parameters.

As a way forward, we explore the impact that data from future surveys like DESI~\cite{DESI:2016fyo} will have by considering BOSS+BAO data with a data covariance matrix rescaled by 16 (see Fig.~\ref{fig:divby16} in Sec.~\ref{sec:divby16}). 
We find that the constraints from Bayesian and frequentist approaches converge to the same interval for $\sigma_8$ as the likelihood dominates over the prior information, suggesting that the issues discussed above will subside with more data. 

Finally, we construct frequentist confidence intervals for five selected $\Lambda$CDM parameters, $\sigma_8$, $h$, $\Omega_m$, $n_s$, $\ln10^{10}A_s$, and compare the constraints from different data sets, including BOSS, eBOSS and \textit{Planck} (see Sec.~\ref{sec:results}). With the profile likelihood, we find that the constraints from BOSS and \textit{Planck} for all five parameters are within $1.4\,\sigma$, finding no indication of a tension. In particular, while the MCMC posterior prefers intervals for $\sigma_8$, which are $1.4\,\sigma$ ($2.5\,\sigma$) lower than the \textit{Planck} value for the WC (EC) EFT parametrization, the intervals from the profile likelihood are only $0.5\,\sigma$ ($0.3\,\sigma$) lower than the \textit{Planck} constraint.
The reduction of the $\sigma$-distances can be mainly attributed to the wide confidence intervals from the profile likelihood, but in the case of $\sigma_8$, also to shifts of the MLE closer to the \textit{Planck} value than the posterior mean. 
In line with previous studies~\cite{Simon:2022lde, Simon:2022csv}, we find that the parameter $\sigma_8$ is most subject to prior effects.
This indicates that the slight ``$\sigma_8$ discrepancy'' seen in the Bayesian results using the EC parametrization is due to the particular choice of priors. On the other hand, although our main profile likelihood analysis makes use of the WC baseline parametrization of the EFTofLSS without priors, we do not expect major changes in our conclusions regarding the state of the $\sigma_8$ tension from resorting to the use of ``priors'' or a different parametrization.

Our results clearly show the advantages and disadvantages of frequentist and Bayesian parameter inference. Since the frequentist inference does not include priors that confine the EFT parameters to the regime intended by the theory, we observe that the data prefers several EFT parameters to take on extreme values, possibly breaking the perturbativeness of the theory. The lack of prior further leads to significantly wider confidence intervals. This loss of constraining power reflects the purely data driven frequentist approach, which is completely agnostic about which model parameters are deemed more likely \textit{a priori}.
On the other hand, the priors in the Bayesian inference are informative and have an impact on the inferred cosmological parameters. This is important since it is not straightforward to define well motivated priors on the EFT parameters, which is reflected in the fact that the WC and EC parametrizations use different standard configurations for the EFT priors. 

Looking towards the future, which will bring more constraining data sets, we can expect these points of discussion to subside as the data will dominate over any subjective preference introduced by the analysis setup. While waiting for better data, our results indicate that the use of frequentist along with Bayesian methods are valuable in order to obtain a fully nuanced view of the data.

\section*{Acknowledgements}

\noindent We thank Pierre Zhang for his comments and insights throughout the project, and Eiichiro Komatsu and Luisa Lucie-Smith for helpful discussions. EBH and LH would like to thank the \textit{Laboratoire Univers \& Particules de Montpellier} for their hospitality, where part of this work was conducted.
We acknowledge computing resources from the Centre for Scientific Computing Aarhus (CSCAA). 
These results have also been made possible thanks to LUPM's cloud computing infrastructure founded by Ocevu labex, and France-Grilles.
E.B.H. and T.T. were supported by a research grant (29337) from VILLUM FONDEN.
This project has received support from the European Union’s Horizon 2020 research and innovation program under the Marie Skodowska-Curie grant agreement No 860881-HIDDeN.
This project has also received funding from the European Research Council (ERC) under the
European Union’s HORIZON-ERC-2022 (Grant agreement No. 101076865).

\appendix

\section{Impact of priors on EFT parameters}\label{ap:A}

\noindent The naturalness of the EFTofLSS framework predicts the EFT nuisance parameters to be of order unity, and too large values of these parameters would break the perturbativeness of the theory~\cite{DAmico:2021ymi}. Thus, the standard WC parametrization described in Sec.~\ref{sec:parameterizations} assigns Gaussian priors on a subset of the nuisance parameters in order to prohibit the non-perturbative regime from influencing the inference.

In principle, such priors could be informed by $N$-body simulations and thereby promoted to likelihoods and interpreted as additional data in the frequentist approach. However, since this is not the case for the above priors, it is statistically not justified to include them in a profile likelihood analysis. 
In the main text, we have illustrated the impact induced by including the priors as likelihoods in the analysis. Here, we repeat this analysis varying the width of the priors. 

\begin{figure}[tb]
    \includegraphics[width=0.5\textwidth]{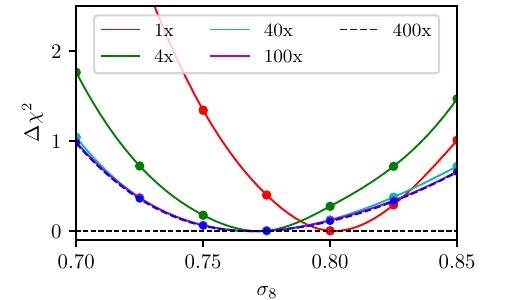}
    \caption{\label{fig:prior_effects} Profile likelihood of $\sigma_8$ under BOSS+BAO data using Gaussian data likelihoods on the EFT parameters, which correspond to the standard WC priors multiplied by different factors indicated by the legend. There is a clear shift in $\sigma_8$ as the prior is widened. In particular, the profiles with widths multiplied by factors of $40$, $100$, and $400$ coincide, indicating that the Gaussian priors reach the limiting case of a flat prior with these large widths. Thus, in our analysis, we model the flat priors on all EFT parameters as the usual priors, but with the widths of the Gaussian priors multiplied by $100$.}
\end{figure}

Flat priors can be modelled as Gaussian priors in the limit that the standard deviations, or \textit{widths}, of the Gaussian priors tend to infinity. Thus, by gradually increasing the width of the standard Gaussian priors, one uncovers the effects of the priors. Fig.~\ref{fig:prior_effects} shows $\sigma_8$ profiles with BOSS+BAO data with the Gaussian priors widths increased by the factor specified in the legend. The red line corresponds to the standard prior configuration of the \texttt{PyBird} likelihood (with $c_4=c_\epsilon^\mathrm{mono} =0$). We observe that the profiles converge to the same shape at large factors, indicating that the Gaussian priors are flat, for all practical purposes, when their widths are increased by factors above $\sim 40$. Accordingly, for convenience purposes in the \texttt{PyBird} code, we model the flat priors on the EFT parameters which have Gaussian priors in the standard configuration by their usual Gaussian priors but with widths multiplied by $100$. 

\begin{figure*}[tb]
    \includegraphics[width=\textwidth]{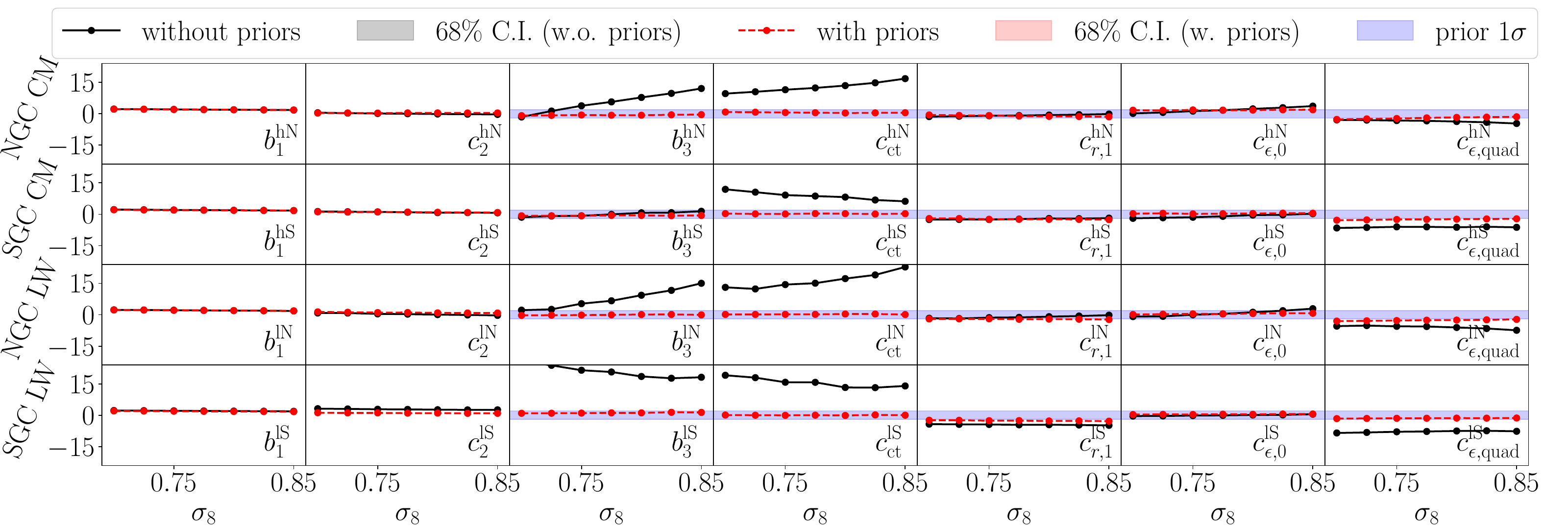}
    \caption{\label{fig:nonmarg_params} Values of the EFT parameters found from optimization at each point in the $\sigma_8$ profile with BOSS+BAO data, with (red) and without (black) the standard WC priors of the \texttt{PyBird} likelihood, described in Sec.~\ref{sec:parameterizations}. The horizontal blue bands illustrate the $1\sigma$ regions of the Gaussian priors. For the parameters without such a band, a flat prior is used ($[0,4]$ for $b_1$ and $[-4,4]$ for $c_2$). The labels CM and LW denote the CMASS and LOWZ galaxy samples, respectively.}
\end{figure*}

The $68\%$ confidence intervals obtained from the $1$x and $100$x widths in the figure are
\begin{eqnarray}
    &\sigma_8 = 0.802 \pm 0.045& \quad \text{(w. prior)} \nonumber \\
    &\sigma_8 = 0.771 \pm 0.075& \quad \text{(no prior)}, \nonumber 
\end{eqnarray}
amounting to a $0.35\sigma$ shift. A similar shift in $\sigma_8$ was found in Ref.~\cite{Simon:2022lde} from an MCMC analysis when increasing the Gaussian priors widths by a factor of $2$. We conclude that the likelihoods imposed on the EFT parameters may influence the constraints when using BOSS data (note, however, that the influence will increase for less constraining data sets and vice-versa).

The disadvantage of not imposing these likelihoods is that one loses control over whether the EFT parameters become too large for the effective field theory description to be appropriate. Thus, the only correct frequentist approach is to let them vary freely and then check explicitly by inspection that they remain of order unity at each point in the profile likelihood. Fig.~\ref{fig:nonmarg_params} shows the values of the EFT nuisance parameters found by optimization at each point in the $\sigma_8$ profile with BOSS+BAO data, both with (red) and without (black) the explicit likelihoods on the EFT parameters. For comparison, the shaded blue region indicates the $1\sigma$ region of the Gaussian prior of the parameters, which have a prior in the standard analysis. We observe that in the case without Gaussian likelihoods mimicking priors, the EFT parameters are \textit{not} of order unity as desired, which can break the perturbative nature of the theory. This result illustrates the conundrum of the priors: either one adopts subjective priors (in a Bayesian framework), which are informative and influence the inferred cosmological parameters, or one works without priors (in a frequentist framework), which leads to extreme values of the nuisance parameters.

\section{Full profile and MCMC results} \label{ap:full_results}
\begin{figure*}[tb]
    \includegraphics[width=\textwidth]{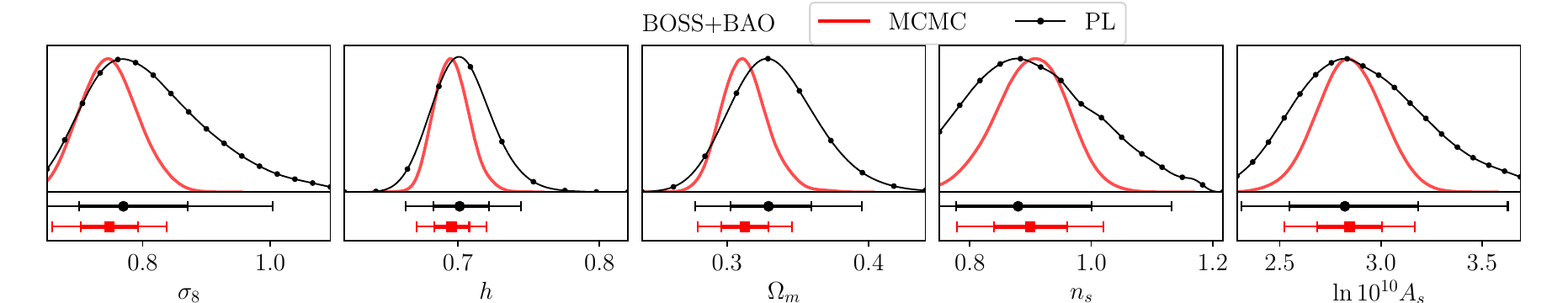}
    \includegraphics[width=\textwidth]{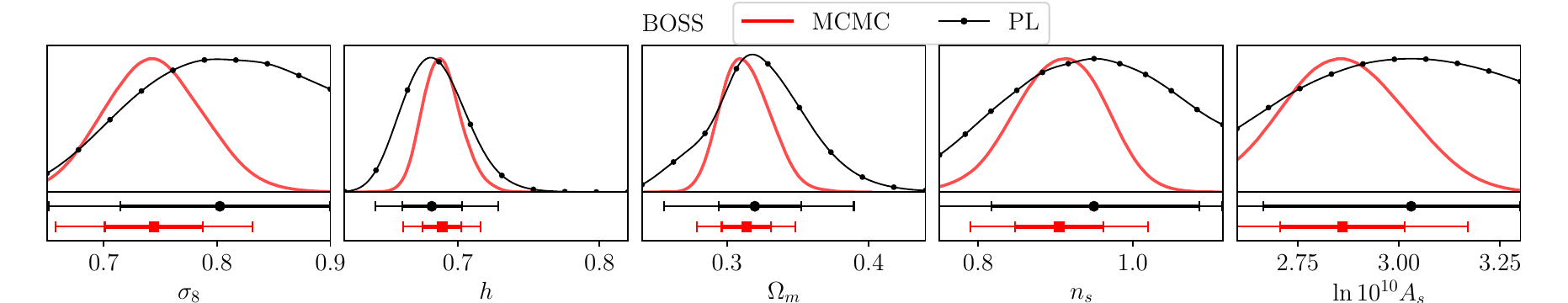}
    \includegraphics[width=\textwidth]{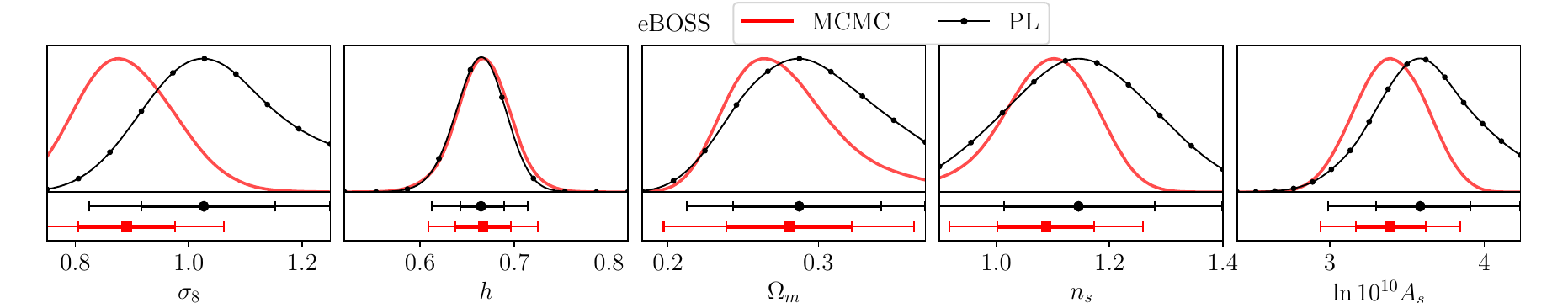}
    \caption{\label{fig:app_comp} Profile likelihoods (black) and one-dimensional marginalized posteriors (red) of the parameters $\sigma_8$, $h$, $\Omega_m$, $n_s$ and $\ln (10^{10} A_s)$ for the data sets BOSS+BAO, BOSS (without BAO post-reconstruction) and eBOSS. The bottom panels show the $68\%$ and $95 \%$ confidence intervals and credible intervals, respectively.}
\end{figure*}

\noindent Fig.~\ref{fig:app_comp} shows the profile likelihoods (black) and one-dimensional marginalized posterior distributions (red) for the BOSS+BAO, BOSS (without BAO post-reconstruction measurements) and eBOSS data sets, derived in this paper. The profile likelihoods are normalized to their MLE. The bottom panels show the $68\%$ and $95 \%$ confidence intervals and credible intervals.

\section{Bestfit parameters}\label{ap:bestfits}
\noindent For the sake of reproducibility, Table~\ref{tab:bestfits} shows the values of the cosmological parameters at the global bestfits found in this work. We note that the bestfits here are simply taken as the point in the profile likelihood with the maximum likelihood; due to the finite sampling of the profile, the bestfit values of these parameters may therefore be slightly inaccurate.

\begin{table}[h]
\begin{tabular}{?Sc|Sc|Sc|Sc|Sc?} 
 \specialrule{.12em}{0em}{0em}
 & BOSS+BAO & BOSS & eBOSS & \textit{Planck} \\ 
\hhline{|=|=|=|=|=|}
$10^2 \omega_b$       & $2.2686$ & $2.2682$ & $2.2674$ & $2.2399$ \\
$\omega_\mathrm{cdm}$ & $0.1391$ & $0.1259$ & $0.1034$ & $0.1198$ \\
$h$                   & $0.7022$ & $0.6838$ & $0.6646$ & $0.6750$ \\
$n_s$                 & $0.8728$ & $0.9270$ & $1.1468$ & $0.9663$ \\
$\ln 10^{10} A_s$     & $2.7925$ & $2.9558$ & $3.5889$ & $3.0442$ \\
\hline 
$\Omega_m$            & $0.3293$ & $0.3191$ & $0.2869$ & $0.3121$ \\
$\sigma_8$            & $0.7699$ & $0.8025$ & $1.0267$ & $0.8100$ \\
\hline 
$\chi^2_\mathrm{min}$ & $138.56$ & $128.38$ & $47.98$ & $1387.07$ \\
 \specialrule{.12em}{0em}{0em}
 \end{tabular}
 \caption{\label{tab:bestfits} Values of cosmological parameters at the global bestfit of the $\Lambda$CDM model under the BOSS+BAO, BOSS, eBOSS and \textit{Planck} data sets, as specified in Sec.~\ref{sec:data}. We stress that the bestfit values here are only approximate due to the finite sampling of the profile likelihoods; a more fair comparison of the constraints is in table~\ref{tab:results}.}
\end{table}

\bibliography{main}

\end{document}